\documentclass[a4paper, 11pt]{amsart}

\usepackage{amsmath, amssymb}
\usepackage[latin1]{inputenc}
\usepackage[english]{babel}
\usepackage[T1]{fontenc}
\usepackage{graphicx}
\usepackage{rotating}
\usepackage{array,multirow}
\newcommand{\R}{\mathbb R}

\def\be#1\ee{\begin{equation}#1\end{equation}}
\newcommand{\fer}[1]{(\ref{#1})}

\newcommand{\bq}{\begin{equation}}
\newcommand{\eq}{\end{equation}}
\newenvironment{equations}{\equation\aligned}{\endaligned\endequation}
\def\bqa{\begin{eqnarray}}
\def\eqa{\end{eqnarray}}

\def\e{\epsilon}

\newcommand{\bd}{\begin{displaymath}}
\newcommand{\ed}{\end{displaymath}}
\newcommand{\ba}{\begin{eqnarray}}
\newcommand{\ea}{\end{eqnarray}}

\def\R{\mathbb{R}}

\theoremstyle{plain}

\title{Social climbing and Amoroso distribution}
\author{Giacomo Dimarco}

\thanks{Department of Mathematics and Informatics of the University of Ferrara
e.mail:   giacomo.dimarco@unife.it 
}
\author{Giuseppe Toscani}

\thanks{Department of Mathematics  of the University of Pavia, and IMATI CNR, Italy.  e.mail:  giuseppe.toscani@unipv.it. 
}

\date{\today}

\begin{document}
\maketitle

\begin{center}\small
\parbox{0.85\textwidth}{

\textbf{Abstract.} 
We introduce a class of  one-dimensional linear kinetic equations of Boltzmann and Fokker--Planck type, describing the dynamics of individuals of a multi-agent society questing for high status in the social hierarchy. At the Boltzmann level, the microscopic variation of the status of agents around a universal  desired target, is  built up introducing as main criterion for the change of status  a suitable value function in the spirit of the prospect theory of Kahneman and Twersky. In the asymptotics of grazing interactions, the solution density of the Boltzmann type kinetic equation is shown to converge towards the solution of a Fokker--Planck type equation with variable coefficients of diffusion and drift, characterized by the mathematical properties of the value function.  The steady states of the statistical distribution of the social status predicted by the Fokker--Planck equations belong to the class of Amoroso distributions with Pareto tails, which correspond to the emergence of a \emph{social elite}.  The details of the microscopic kinetic interaction allow to clarify the meaning of the various parameters characterizing the  resulting equilibrium. Numerical results then show that the steady state of the underlying kinetic equation is close to Amoroso distribution even in an intermediate regime in which interactions are not grazing.


\medskip

}

\medskip


\end{center}
\section{Introduction}

More than one century ago, the economist Vilfredo Pareto \cite{Par1,Par2} observed that human societies tend to organize in a hierarchical manner, with the emergence of \emph{social elites}.  He further noticed that social mobility in this hierarchical state appears to be higher in the middle classes than in the upper and lower part of the hierarchy. 
The correctness of Pareto's social analysis has been recently investigated in a number of studies dealing with the modeling of the dynamics of social networks in which individuals quest for high status in the social hierarchy \cite{BDL,KTessone,KTZ,ZTY}. There, the social status  of individuals is usually ranked according to a certain measure strongly related to centrality in the society. Since individuals in top ranked positions enjoy social advantages compared to those with low rankings, individuals are motivated to climb the social ladder to reach a better rank \cite{Cha}.

These models captured at a numerical level some key ingredients enough to reproduce the main characteristics predicted by the sociological analysis of Pareto \cite{Par1,Par2}. In particular the numerical simulations of the dynamics driven by the models introduced in  \cite{BDL,Cha} led to conclude that, in accord with Pareto's observations, the hierarchical state reached by individuals in the social network looking for high status is very stable.

By contrast with the numerical simulations that form the core of most previous studies, we show that the analytic techniques of statistical mechanics are ideally suited to the study of the phenomenon of social climbing, and can shed much light on its structure and behavior. 
In particular, by resorting to the well-established mathematical tools of kinetic theory of multi-agent systems \cite{PT13},  we pursue an almost entirely analytic approach. 
Starting from a suitable description of the microscopic behavior of agents, we build a class of  one-dimensional linear kinetic equations of Boltzmann and Fokker--Planck type, suitable to describe the dynamics of individuals of a multi-agent society tending for high ranking in the social hierarchy, in which both the properties of the transient and of the steady state are explicitly computable.  This approach is coherent with the classical kinetic theory of rarefied gases, where the formation of a Maxwellian equilibrium in the spatially uniform Boltzmann equation is closely related to the microscopic details of the binary collisions between molecules \cite{Cer,GT-ec}. 

The methods and results of this paper are part of the numerous studies dealing with the modeling of social and economic phenomena in a multi-agent system. In economics, this modeling attempted to justify the genesis of the formation of Pareto curves in wealth distribution of western countries \cite{ChaCha00,CCM,ChChSt05,CoPaTo05,DY00,GSV,SGD},  and to share light on the reasons behind opinion formation \cite{BN2,BN3,BN1,BeDe,Bou,Bou1,Bou2,CDT,DMPW,GGS,GM,Gal,GZ,SW,To1}.

Due to the human nature of social phenomena, these investigations naturally included specific behavioral aspects of agents in the modeling. In the field of collisional kinetic theory a pioneering approach has been proposed in   \cite{MD} to model price formation of a good in a multi-agent market. The kinetic model of Boltzmann type describes the dynamics of  two different trader populations, playing different rules of trading, including the possibility for agents to move from one population to other. The kinetic description, inspired by the well-known Lux--Marchesi model \cite{LMa,LMb} (cf. also  \cite{LLS,LLSb}) included in the mechanism of trading  the opinion of traders \cite{To1},  and behavioral components of the agents, like risk's perception. This last component has been done, by resorting to the prospect theory by Kahneman and Twersky \cite{KT,KT1}, in terms of a suitable \emph{value function}. 

The analysis of   \cite{MD} enlightened the importance of modeling kinetic interactions by taking into account aspects of human behavior \cite{BHT,BKS,BCKS,BGO},   as pioneered by Zipf in  \cite{Zipf} (a recent collection of contributions on this topic can be found in two special Issues of this journal \cite{BB1,BB2}). The possible connections between the kinetic modeling of human phenomena and their description in terms of value functions, has been recently developed in  \cite{GT17}, where the mathematical translation of this relationship  justified at a microscopic level first the mechanism of formation of the service time distribution in a call center, and subsequently a number of other social phenomena which lead to a stationary state in the form of a lognormal distribution \cite{GT18}.

The leading idea in  \cite{GT17}  is based on a general principle which can be easily verified in a number of social activities of agents in which one identifies the possibility of a certain addiction \cite{DT,To2}.

The forthcoming kinetic modeling follows along the lines of the recent approaches of  \cite{DT,GT17,GT18}, where a relevant number of phenomena involving measurable quantities of a population and fitting suitable probability distributions (in particular log-normal distribution) was shown to be  consequence of a certain human behavior. The list of social phenomena to which the kinetic description furnishes a convincing explanation of the reasons behind the formation of a certain statistical profile is remarkable. Furthermore, the kinetic approach of  \cite{GT17,GT18}, recently extended to addiction phenomena \cite{DT,To2}, gives a unified view to various social aspects, previously treated resorting to different methods. The list include the distribution of body weight \cite{BC}, women's age at first marriage \cite{Pre},  drivers behavior \cite{JJ},  the level of alcohol consumption \cite{Keh,Reh},  or, from the economic world, the level of consumption in a western society \cite{BBL}, the size of cities \cite{BRS}, the length of call-center service times \cite{Brown}.

In agreement with the modeling assumption of  \cite{GT18}, the microscopic agent's behavior is expressed by assuming that the agents change  their state aiming to approach an optimal target, and that this change requires an asymmetric effort, depending on whether the current state is above or below the optimal one. On the basis of the prospect theory by Kahneman and Tversky \cite{KT,KT1},  the elementary changes of state of agents depend on a suitable value function expressing this asymmetry in the whole range of possible values. In the situation leading to the log-normal or other rapidly decaying probability densities, the asymmetry of the value function translates at the mathematical level a fundamental property: it is easier to reach the optimal target starting from below,  than to approach it from above.

To clarify this idea through a simple example, an asymmetry of this type is present when looking at body weight of individuals characterized by common age and sex, where the desired target can be identified in the mean \emph{ideal weight}. It is clearly easier to reach the ideal weight starting from lower values, since this implies no restrictions on eating, while it is hard to reduce the weight when above the ideal target, since this requires to respect a certain diet. 

In analogy with the previous phenomena, the analysis of social climbing leads to conclude that the microscopic dynamics still contains a marked asymmetry, which, however, for low values  of the social rank,  works in the opposite direction. Indeed, while it is clear that  agents tend as before to improve their social status, trying to approach an optimal target that can be identified in a certain level of well-being, it is usually hard to exit from the very low zone, since this implies to possess, in addition to personal skills and competences,  a certain amount of financial resources, possibility of access to good schools,  and other facilities.

With respect to the choice made in  \cite{GT18,To2}, this asymmetry will be expressed through a new class of value functions, still obtained  from the classical ones proposed by Kahneman and Tversky \cite{KT,KT1}, which now express Pareto's analysis about social mobility: the microscopic variations are higher for agents in the middle class than for agents in the upper and lower part of the hierarchy. The new value functions are used to built the microscopic interaction suitable to describe the elementary variation of the social rank, and to study in this way the phenomenon through the variation in time of the density $f= f(w,t)$ of agents in the system with social rank  measured by $w >0$, at time $t\ge 0$.
 
As usual in the kinetic setting \cite{FPTT}, the change of $f$ due microscopic interactions generates a linear kinetic equation of Boltzmann type for the density of agents, that will be subsequently studied in the asymptotic regime of \emph{grazing} effects \cite{Vi}. 
The  \emph{grazing} regime describes the situation in which a single interaction produces only a very small change of the variable $w$. In this regime, the variation of density of the social rank of the agent's system is driven by a partial differential equation of Fokker--Planck type. If we denote by $g = g(w,t)$ the density of agents which have a  social rank  equal to $w$ at time $t\ge 0$, this density is solution of a linear Fokker--Planck equation with variable coefficients of diffusion and drift, given by
\begin{equation}\label{FPori}
 \frac{\partial g(w,t)}{\partial t} = \left\{\frac \sigma 2 \frac{\partial^2 }{\partial w^2}
 \left(w^{2+\delta} g(w,t)\right )+ \frac \mu{2}
 \frac{\partial}{\partial w}\left[ \frac 1\delta \left(1- \left(\frac{\bar w_L}w \right)^\delta \right)w^{1+\delta} g(w,t)\right]\right\}.
 \end{equation}
In \fer{FPori}  $\bar w_L$ represents the target value of social rank that agents tend to reach,  while $\sigma, \mu$ and $ \delta$  are positive constants  closely related to the typical quantities of the phenomenon under study, satisfying further bounds imposed by the physical conditions on the value function in the microscopic interaction. In particular $0 < \delta \le 1$. The equilibrium density of the Fokker--Planck equation \fer{FPori} is given by the Amoroso-type distribution \cite{Amo}
 \be\label{equili}
g_\infty(w) = g_\infty(\bar w_L)\left( \frac{\bar w_L}w \right)^{2 + \delta +\gamma/\delta}  \exp\left\{ - \frac \gamma{\delta^2}\left( \left( \frac{\bar w_L}w \right)^\delta -1 \right)\right\}.
 \ee 
In \fer{equili} $\gamma= \mu/\sigma$, where $\sigma$ and $\mu$ are the coefficients of the diffusion (respectively of the drift) terms of the Fokker--Planck equation \fer{FPori}. This asymptotic procedure was used in  \cite{CPP,DMTb} for  a kinetic model for the distribution of wealth in a simple market economy subject to microscopic binary trades in presence of risk, showing formation of steady states with Pareto tails, in  \cite{TBD} on kinetic equations for price formation, and in  \cite{To1} in the context of opinion formation in presence of self-thinking. A general view about this asymptotic passage from  kinetic equations based on general interactions  towards Fokker--Planck type equations can be found in  \cite{FPTT}, where also an exhaustive discussion about the large-time behavior of the solution to equations \fer{FPori} is presented. Other relationships of this asymptotic procedure with the classical problem of the \emph{grazing collision limit} of the Boltzmann equation in kinetic theory of rarefied gases have been recently enlightened in  \cite{GT17}. 
 
Assuming that the steady state distribution \fer{equili} is taken of unit mass, the equilibrium state is a probability density belonging to the class
 \be\label{equi}
f_\infty(w;\theta, \alpha,\beta) =  \frac 1{\Gamma(\alpha)}\left| \frac \beta\theta\right|\left(\frac w\theta\right)^{\alpha\beta -1}
\exp\left\{ - \left( w/\theta\right)^\beta\right\},
 \ee 
for non-negative values of $w$ and positive values of the parameters $\alpha, \theta$ and $\beta\not= 0$. When $\beta >0$, the function \fer{equi} was considered as a generalization of the Gamma distribution by Stacy \cite{Sta}, and includes the familiar Gamma, Chi, Chi-squared, exponential and Weibull densities  as special cases. Generalized Gamma distributions, also known as Amoroso and Stacy-Mihram distributions \cite{Amo,JKB},  are widespread in physical and biological sciences \cite{JHMG,Li,Lie}, as well as in the field of social sciences \cite{box,DT}.

Note that the density \fer{equili} corresponds to the choice of the negative value  $\beta = -\delta$ in \fer{equi}. Negative values of $\beta$ lead to distributions with only a limited number of moments bounded, and represent, among others, a careful approximation of the statistical distribution of wealth,  in accord with Pareto's discovery of fat tails in this case \cite{Par}.  In particular, if $\delta =1$ the function \fer{equili}  coincides with an inverse Gamma distribution. 

The choice of $\beta = -1$ establishes a strong connection between the steady distributions of social rank, and, respectively,  of wealth. As a matter of fact, this connection is evident if we consider that the social status  of individuals is closely related to their wealths. However, the mechanism of wealth formation in a multi-agent system has been modeled at a kinetic level in terms of binary trades.
Beside the kinetic models of Boltzmann type introduced in recent years to enlighten the formation of an unequal distribution of wealth among trading agents \cite{CCCC,PT13}, a Fokker--Planck type equation assumed a leading role. This equation, which describes the time-evolution of the density $f(t,w)$ of a system of agents with personal wealth $w\ge0$ at time $t \ge 0$ reads
  \be\label{FP2c}
 \frac{\partial f(w, t)}{\partial t} = \frac \sigma{2}\frac{\partial^2 }
 {\partial w^2}\left( w^2 f(w,t)\right) + \frac\lambda{2} \frac{\partial }{\partial w}\left(
 (w-m) f(w,t)\right).
 \ee
In \fer{FP2c},  $\sigma, \lambda$, and $m$ denote  positive constants related to essential properties of the trade rules of the agents.  

The Fokker--Planck equation \fer{FP2c} has been first obtained by Bouchaud and M\'ezard \cite{BM} through a mean field limit procedure applied to a stochastic dynamical equation for the wealth density.  Then, the same equation was derived in   \cite{CoPaTo05} by resorting to an asymptotic procedure applied to a Boltzmann-type kinetic model for binary trading in presence of risk. Note that the steady state of equation \fer{FP2c} is an inverse Gamma density. 

The Boltzmann-type equation leading to \fer{FP2c}, was assumed to satisfy the strong hypothesis of Maxwellian molecules. This choice  has been critically revisited in  \cite{FPTT1}, to give a more coherent (from the economical point of view) interaction kernel. There, the choice of a variable kernel gave in the grazing limit a Fokker--Planck equation different from \fer{FP2c}, that while possessing an inverse Gamma density as steady state, allowed to prove that the solution is converging to equilibrium at exponential rate, a result that is missing for the solution to \fer{FP2c}. The new Fokker--Planck equation considered in  \cite{FPTT1} is
  \be\label{FP-new}
 \frac{\partial f(w, t)}{\partial t} = \frac \sigma{2}\frac{\partial^2 }
 {\partial w^{2}}\left( w^{2+\nu} f(w,t)\right) + \frac\lambda{2} \frac{\partial }{\partial w}\left(
 (w-m)\, w^\nu f(w,t)\right),
 \ee
where  $0 < \nu \le 1 $ is a constant parameter related to the intensity of the frequencies described by the kernel in the Boltzmann equation. The Fokker--Planck equation \fer{FP-new} interpolates between equation \fer{FP2c} ($\nu =0$) and equation \fer{FPori} ($\nu=\delta =1$). 

The forthcoming kinetic modeling  is not restricted to the sociological aspect of the formation of elites, but it can be easily adapted to describe the distribution of knowledge in a society \cite{GT-ec,PT4}, where the elites in this case can be identified for example in the members of the scientific academies, or in the description of social climbing in  sports with a large participation, such as the football game, where the elites correspond to the footballers playing in the best clubs of a country. This last example is particularly interesting since one can be easily access data, and furthermore allows to understand in a clear way the main assumptions leading to the general kinetic description. 

In details, in the forthcoming Section \ref{football} we will present on the example of social climbing in football activity, the main \emph{universal} assumptions  modeling the microscopic kinetic interaction, and  the consequent  interaction kernel. Then, in Section \ref{wealth}, we discuss from a different point of view the classical description of the trading interaction in the linear kinetic model for wealth distribution leading to the Fokker--Planck equation \fer{FP2c}, which is now rewritten in terms of the value function formulation adopted in  \cite{GT17}. This preliminary discussion allows to clarify  the main differences between the kinetic models for wealth distribution based on linear trades, and the present one for social climbing, mainly motivated by social reasons. This will be presented in  Section \ref{model}, where we describe a multi-agent system, in which agents can be characterized in terms of  their social rank, measured in terms of a certain unit, and subject to microscopic interactions which contain the microscopic rate of change of the value of their rank, according to the previous general principle. 
The relevant mechanism of the microscopic interaction is indeed based on a suitable value function, in the spirit of the analysis of Kahneman and Twersky \cite{KT,KT1}, which reproduces at best the asymmetries present in the human behavior in this situation.
 Then in Section \ref{quasi} we will show that in a suitable asymptotic procedure (hereafter called \emph{grazing}  limit) the solution to the kinetic model tends towards the solution of the Fokker-Planck type equation \fer{FPori}.

Once the  Fokker--Planck equation \fer{FPori} has been derived, some numerical examples  will be collected in Section \ref{numerics}, together with a detailed explanation  of the relevant mechanism which leads to the typical microscopic interaction in terms of the value function. The numerical experiments will put in evidence that the time behavior of both the kinetic model and its Fokker--Planck asymptotics is very similar, and that the steady states are very close each other even in for  moderate values of the grazing parameter. 

\section{The example of football activity}\label{football}

\begin{figure}\centering
	{\includegraphics[width=12.5cm]{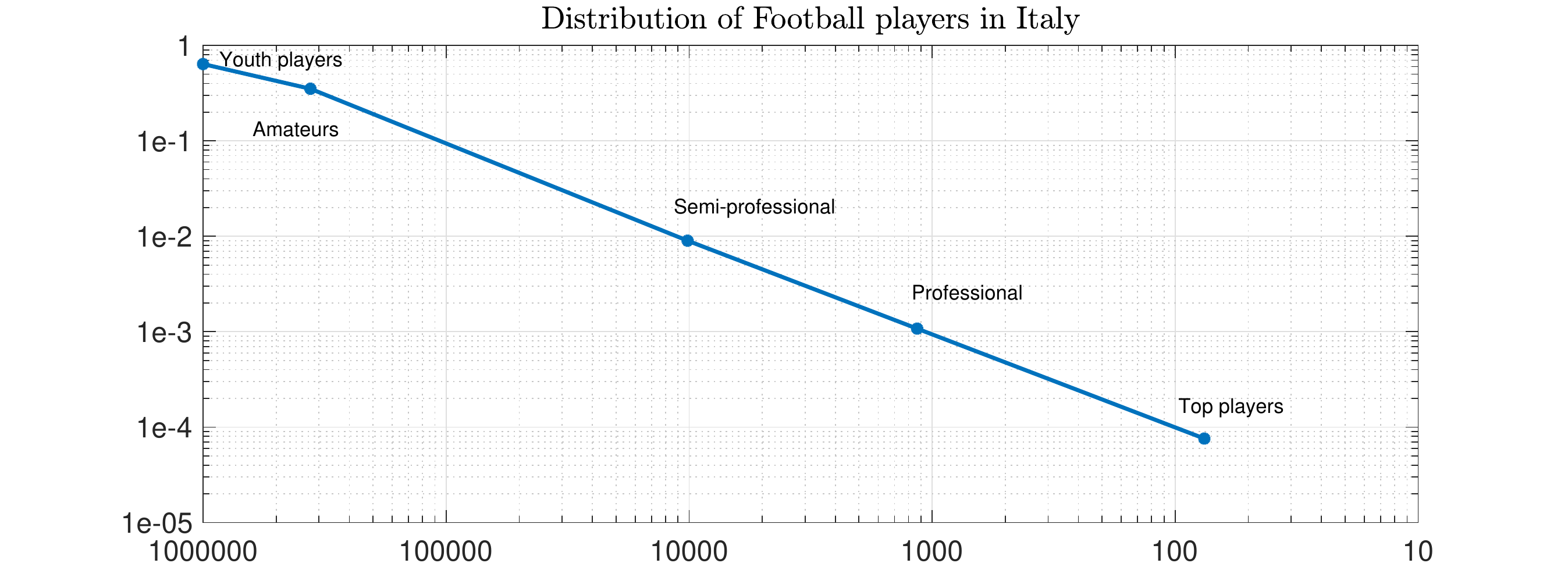}}
	\caption{Distribution of football players in Italy from youth schools to top clubs, 2017-2018 census.}\label{fig0}
\end{figure}

The social hierarchy represents a fundamental aspect of life observed in many different animal species, including insects, mammals and birds. Depending on the situation, a hierarchy may be established in different manners but it often results in a ranking of the animals in a group and, its importance is such that can influence the quality of life of the entire group \cite{Sa}.  It is clear that human beings are far from being unaffected from this phenomenon. On the contrary, in human societies, one can observe in everyday life, an enormous number of examples which could be possibly, in last instance, be related to a construction of a hierarchical structure. At this subject, Figure \ref{fig0} shows the distribution of football players in Italy starting from the young player academies up to professionalism. In this plot, Top players are considered those players belonging to the top five Italian clubs. The image is produced using the data reported in Table \ref{table1} extrapolated from the Annual Report of the National Football Italian Association (FIGC)\footnote{\label{note1}https://www.figc.it/it/federazione/federazione-trasparente/reportcalcio/}.This table shows the number of Italians playing per categories together with the inverse cumulated number of people belonging to each class. The first level regards children starting to play football, the second one people that play independently on the age at different amateur levels, while the third level contains the young professionals (referred to as semi-professionals). In the fourth class, we extrapolated the estimated number of professional players coming from the Italian sport Academies. This datum has been obtained eliminating the number of players from Foreign countries from the total number of registered professionals playing in the Italian leagues at the moment of the survey. Finally, the last level (the \emph{football elite}) gives the number of estimated players coming from the Italian Academy sector arriving in the top five Italian clubs. This last information has been extrapolated considering the average number of players in each club and discarding the number of foreign practitioners. The Figure is obtained by reporting the relative number of players belonging to each category versus the inverse cumulative distribution, which is intended as the measure of the social position in the football society, in logarithmic scale. One can observe that the empirical inverse cumulated distribution exhibits tails. A similar Figure, concerning the distribution of school knowledge in Italy, has been obtained in  \cite{GT-ec} using the data relative to the 2011 census. In this case, the \emph{knowledge elite}  was represented by members of the historical scientific academies.

\begin{table}
	\begin{center}
		\begin{tabular}{lll|ll}
	Level & Italians & (\%) & Cumulated values & (\%)\\ 
	\hline
	Youth players & 673555 & 63.6\% &1054791 & 100\% \\
	Amateurs &370540 & 35.1\% &381236 &36.14\%\\
	Young professional	& 9480 & 0.9\% &10696 &1.014\%\\
	Professional &1136 &0.108\% & 1216 & 0.1152\% \\
	Top players & 80 & 0.007\% &80 & 0.007\%\\
	\hline
	Totals &1054791 & 100& &\\
	\end{tabular}\label{table1}
\end{center}
	\caption[]{Distribution of the players for level of activity. Data from FIGC$^{\ref{note1}}$ }
\end{table}
The example of climbing in football activity can serve as a prototype case to understand and justify the main rules of both the personal and social behavior of players characterizing the kinetic model described in the rest of the paper. Inherently to the football activity, progress into the social ladder can be undoubtedly identified in playing for the best clubs of the first Italian league. Referring to players at any level to as agents (borrowing this term from classical microscopic modeling approaches), it is reasonable to assume that improving in the practice of football (or any sport in general) can be reached only through a very large number of small steps: everyday trainings routine for example. In this context, an agent tries to reach an optimal target identified as the level in which a young football player aims to play when becoming an adult. In this long process, it is also clear that while it is hard to improve at the beginning, progresses become faster as the young players grow and learn how to practice. This point of view shares similarities with the common perception that people belonging to low social classes have low chances to improve their status for lack of means and thus often remain in the so-called bottom of the hierarchy. In the same way, animals in dominant hierarchies do not consider themselves sufficiently strong to attempt any climb and so they simply give up. Coming back to the football example, it is considered usually hard to overcome the low skills level zone in football, since this implies to possess, in addition to inbred skills, a certain  possibility of access to good instructors and facilities. Trying to make quantitative the previous observations, we will in the next part of the paper introduce a variable $w$ identifying the social status in a society (the football skills in our example). In the same way, as it will be made clear next, we will introduce a value $\bar w$ identifying the limit below which an agent does not expect to be able to climb the social ladder. This point will characterize a change in the way in which the agent will be able to adjust its social status: passed this value his improvements will be faster. In the same way, many football players consider reaching a certain level in the league enough satisfactory for them. This is also analogous to what happens to people satisfied of obtaining a certain job not questing for changes. In order to take into account this second fact, we will define then with $\bar w_L > \bar w$ this perceived level which is assumed, at first glance, equal for all agents. The other ingredients which completely defines the dynamic under study are the will of football players to improve, as well as, a certain dose of uncertainty depending on external factors which cannot in general be controlled. All these aspect will be condensed in a so-called value function $\Psi$ in the sense of Kahneman and Twersky.\cite{KT} Finally, regarding players which already reached the professional level, we expect for them as natural to be difficult to descend back to lower levels apart if rare circumstances happen. In the next Section, we will discuss the analogies of the proposed model with more classical models describing wealth distribution \cite{PT13}.

\section{Learning from kinetic theory of wealth distribution}
\label{wealth}
The brief discussion of Section \ref{football} illustrates the main reasons which can justify the profile of Figure \ref{fig0}.  These reasons need now to be translated at a mathematical level in a model, in order to verify if, in consequence of these rules,  this steady behavior appears.

Among other approaches, the description of  social phenomena in a multi-agent system can be successfully obtained by resorting to statistical physics, and,  in particular, to methods borrowed from kinetic theory of rarefied gases. The main goal of the mathematical modeling is to construct master equations of Boltzmann type, usually referred to as kinetic equations, describing the time-evolution of some characteristic of the agents, like wealth, opinion, knowledge, or, as in the case treated in this paper, of agent's ranking in the social ladder \cite{CaceresToscani2007,CCCC,DT,NPT,PT13,SC}.

The building block of kinetic theory is represented by the details of microscopic interactions, which, similarly to binary interactions between particles velocities in the classical kinetic theory of rarefied gases, describe the variation law of the selected agent's trait.  Then, the microscopic law of variation of the number density consequent to the (fixed-in-time) way of interaction, is able to capture both the time evolution and the steady profile. 

In economics, the microscopic interactions attempted to simulate the trading activity, aiming to justify in this way the formation of Pareto curves in wealth distribution of western countries \cite{ChaCha00,CCM,ChChSt05,CoPaTo05,DY00,GSV,SGD}. The qualitative results obtained in  \cite{DMTb,MaTo} then showed the possibility that Pareto tails could be obtained in consequence of the linear binary trades between agents introduced in  \cite{CoPaTo05}. For a better understanding of the reasons leading in general to the formation of Pareto tails, 
in the rest of this section we will give a new insight to this linear trading.
 
We consider a system of agents characterized by their wealths. To fix notations, the amount of wealth will be denoted by $v \ge 0$, and measured with respect to some unit. 
As usual, the population  is considered homogeneous with respect to the way of trading.   
In addition, it is assumed that agents are indistinguishable \cite{PT13}. This means that  at any instant of time $t\ge 0$ agents in the system are completely characterized by the amount  of their wealths. 
Consequently, the statistical distribution of wealth of the agents system will be fully characterized by the unknown density  $f = f(v,t)$, of the wealth $v\in \R_+$ and the time $t\ge 0$. 

The precise meaning of the density $f$ is the following. Given the system of traders, and given an interval or a more complex sub-domain $D  \subseteq \R_+$, the integral
\[
\int_D f(v,t)\, dv
\]
measures the amount of traders which  are characterized by a  wealth $v \in D$  at time $t \ge 0$. It is assumed that the density function is normalized to one, that is for all $t\ge 0$
\be\label{uno}
\int_{\R_+} f(v,t)\, dv = 1.
\ee
The change in time of the density is due to the fact that agents of the system are subject to trades, and continuously upgrade their amounts of wealth $w$  at each trade.  To maintain the connection with classical kinetic theory of rarefied gases, it is usual to define \emph{elementary interaction} a single upgrade of the quantity $v$. 

Following   \cite{FPTT1}, we will limit ourselves to consider a linear interaction, which while taking into account all the trading aspects of the original nonlinear model considered in   \cite{CoPaTo05}, and giving in the asymptotics of grazing collisions the same Fokker--Planck equation, it is easier to handle. According to the binary trade introduced in   \cite{CoPaTo05}  the elementary change of wealth $v \in \R_+$  of an agent of the system trading with the market is the result of three different contributes
 \be\label{lin}
 v^* = (1-\e\lambda)v +  \e\lambda \bar v  + \eta_\e\, v.
 \ee  
In \fer{lin} $\lambda$ is a positive constant, while $\e \ll 1$ is a small parameter that adjusts the intensity of the exchange, and guarantees that $\e\lambda\ll 1$. The first term in \fer{lin} measures the wealth  that remains in the hands of the trader  who entered into the trading market with a (small) percentage  $\e\lambda v$ of his wealth. The constant $\e\lambda$ quantifies the saving propensity of the agent, namely the human perception that it results quite dangerous to trade the whole amount of wealth in a single interaction. The second term represents the amount of wealth the trader receives from the market as result of the trading activity.  Here $\bar v >0$ is sampled by a certain distribution $\mathcal E$ which describes the (independent of time) distribution of wealth in the market. Note that in principle the constant in front of the wealth $\bar v$ could be different from $\lambda$, say $\xi$. However,  the  choice $\lambda \not=\xi$ will not introduce essential differences in the subsequent discussion. Finally, the last term takes into account the risks connected to the trading activity. The uncertainty of the trading result is represented in terms of a random variable  $\eta_\e$,   centered and with finite variance $\e\sigma \ll1$, which in general  is assumed such that $ \eta_\e \ge -1 + \e\lambda$, to ensure that even in a risky trading market, the post trading wealth remains non negative.  In  \cite{MaTo} it is further assumed that the random variable $\eta_\e$ takes values on a bounded set, that is $-1 + \e\lambda \le \eta_\e \le \e\lambda^* < +\infty$. This condition is coherent with the trade modeling, and corresponds to put a bound from above at the possible random gain that a trader can have in a single interaction.

Since $\bar v >0$, the elementary interaction \fer{lin} is a particular case of interactions in which the value  $v$ of the wealth can be modified by two quantities which describe the predictable, and, respectively, the unpredictable behavior of the outcome. The general form of these interactions can be written in the form
\be\label{lin2}
 v^* = v - \Psi^\e\left( \frac v{\bar v} \right) v  + \eta_\e\, v.
 \ee 
Using the form \fer{lin2} for the interaction \fer{lin} we obtain that, for $s = v/\bar v \ge 0$
 \be\label{val}
  \Psi^\e(s) = \e\lambda\left(1 -\frac 1s \right).
 \ee
Note that $\Psi^\e(s)$ is a dimensionless increasing concave function, which ranges from $-\infty$ to $\e\lambda$, equal to zero at the point $s=1$. This point is usually referred as \emph{reference point}. The role played by the function $\Psi^\e(\cdot)$ in the interaction is clear. Since $\Psi^\e(\bar v/v)$ is negative when $v < \bar v$, and positive when $v > \bar v$ the interaction will increase the value of the wealth in the former case, while it will decrease the value in the latter. Hence, in absence of randomness, the interaction  will move the wealth $v$ towards the value $\bar v$. 

According to the prospect theory of Kahneman and Twersky \cite{KT}, the function $\Psi^\e$ satisfies most of the properties of a \emph{value function}.
The notion of value function was originally related to various situations concerned with decision under risk. Kahneman and Twersky in  \cite{KT} identified the main properties characterizing a value function $\Phi(s)$, where $s \ge0$, in a certain behavior around the reference point $s=1$, expressed by the conditions
 \be\label{ccd}
 - \Phi\left(1-\Delta s \right) > \Phi\left(1+ \Delta s \right),
 \ee
 and 
 \be\label{cce}
 \Phi'\left(1+\Delta s \right)< \Phi'\left(1-\Delta s \right). 
 \ee
 where $\Delta s >0$ is such that $1- \Delta s \ge 0$.
 These properties are well defined for deviations from the reference point $s = 1$, and imply
that the value function below the reference point is steeper than the value function above it.
In other words, a value function is characterized by a certain asymmetry with respect to the
reference point $s = 1$. This asymmetry has a precise meaning. Given  two agents starting at the same distance $\Delta s$ from the reference value $s=1$ from below and above, it will be easier for the agent starting below to move closer to the reference value, than for the agent starting above.The function $\Psi^\e(s)$ in \fer{val} satisfies properties \fer{ccd} and \fer{cce}. A third property follows by considering its behavior as $\e \to 0$. Since
\be\label{dde}
 \lim_{\e \to 0} \frac{\Psi^\e(s)}\e = \lambda\left(1 -\frac 1s \right).
 \ee
 the scaled function $\Psi^\e(s)/\e$ is inversely proportional to $s$.
 
 The main result in  \cite{CoPaTo05} was to show that, in the limit $\e \to 0$, the solution to the kinetic equation of Boltzmann type converges towards the solution to the Fokker--Planck equation \fer{FP2c}, with a steady state in the form of an inverse Gamma density. The inverse Gamma density is obtained as the distribution of the random variable $1/X$, where $X$ is Gamma distributed. This suggests that a Fokker--Planck equation with a steady state in the form of a Gamma density is obtained by looking at the interaction \fer{lin2}, characterized by a value function  that gives the variation of $z=1/v$. 
 To this aim, let us study of the variation of $z= 1/v$.  In absence of risk, the interaction \fer{lin2},  rewritten in terms of $z$ reads
 \be\label{inv}
 \frac 1{z^*} = \frac 1z - \Psi^\e\left( \frac {\bar z}z \right)\frac 1z.
 \ee
Hence, starting from \fer{inv}, simple computations  show that the elementary variation of $z$ is given by
 \be\label{true}
  z^* = z - \Phi^\e\left( \frac z{\bar z} \right) z.
    \ee
In \fer{true} the function $\Phi^\e$, for $s \ge 0$ is given by
 \be\label{val2}
  \Phi^\e(s) = \e\lambda \frac{s -1}{\e\lambda(s-1)+ 1}.
 \ee
Note that the condition $\e \lambda <1$ implies the positivity of the denominator. The function
 $\Phi^\e(s)$ satisfies the bounds
 \[
 -\frac{\e\lambda}{1- \e\lambda} \le \Phi(s) \le 1.
 \] 
Hence, at difference with interaction \fer{lin}, the elementary variation of $z =1/v$ is driven by a function  $\Phi^\e$, bounded from below and above. 

It is interesting to remark that $\Phi^\e$ still satisfies  properties \fer{ccd} and \fer{cce}, typical of a value function. Moreover, as it happens for $\Psi^\e$, the quotient between $\Phi^\e$ and $\e$ remains well-defined in the limit $\e \to 0$, and 
\be\label{ddf}
 \lim_{\e \to 0} \frac{\Phi^\e(s)}\e = \lambda\left(s-1 \right),
 \ee
so that the scaled function is proportional to $s$. 
Motivated by the study of the statistical distribution of alcohol consumption, in  \cite{DT} a  value function suitable to describe the elementary interaction was identified in 
\be\label{vvd}
 \Phi_1^\e(s) = \mu \frac{e^{\e(s -1)}-1}{e^{\e(s -1)}+1 } , \quad  s \ge 0,
 \ee
where the constants $0 <\mu <1$ and $\e >0$ characterize the intensity of the interaction. This function satisfies  properties \fer{ccd} and \fer{cce}, and 
\be\label{ddg}
 \lim_{\e \to 0} \frac{\Phi^\e_1(s)}\e = \mu(s-1),
 \ee
namely the same scaling property of the function \fer{val2}. As proven in  \cite{DT},  in presence of a elementary interaction of type \fer{lin2}, with value function \fer{vvd}, the kinetic model   leads indeed to a statistical distribution in the form of a Gamma distribution.  

This suggests that value functions satisfying the limit property \fer{ddg} generate statistical distribution with thin tails, while fat tails are generated by value functions satisfying a scaling property like \fer{dde}. We will make use of this observation in the next Section.

To conclude this discussion, it is important to remark that interaction \fer{lin}, while leading to a kinetic equation with a steady state that can have fat tails in presence of high risk \cite{MaTo}, it is based on the value function \fer{val} which gives a big increment to a small wealth $v\ll \bar v$.
This behavior is clearly in contrast with the common belief that to increase the personal wealth by trading, starting from a low value of wealth,  is usually much more difficult than to increase it starting from a high level. As recently discussed in  \cite{FPTT1}, this apparent inconsistency of the model can be corrected by introducing a variable frequency of interactions that penalizes interactions in which the wealth $v$ put into the trade is small. We will be back to this question later on.

\section{The kinetic description of social climbing}
\label{model}
\subsection{Modeling the elementary interaction}

In this Section, we will deal with the mathematical modeling of the elementary interaction in the social climbing, treasuring the  discussions of Section \ref{football}  about climbing in football activity, and that of Section \ref{wealth} about the kinetic modeling of wealth distribution.
As before, we assume that the behavior of the population with respect to the climbing of social ladder is homogeneous. This homogeneity assumption is clearly quite strong in general, since it requires at least to restrict the population with respect to some characteristics, like age, sex and degree of education. 

Once the homogeneity assumption is satisfied, the agent's state at any instant of time $t\ge 0$ is completely characterized by the value $w \ge0$ of the social rank  occupied in the society.  We assume that this value can be measured in terms of some reasonable unit. If  the relationship between social rank and personal wealth is identified as substantial, it is possible to measure the value $w$ with the unit of the wealth of agents. However, depending on the homogeneous group of the society we are considering, other choices can be equally possible. 
The unknown is the density (or distribution function) $f = f(w, t)$, where $w\in \R_+$ and the time $t\ge 0$, and the target is to study the statistical features of the subsequent steady state.

Since agents in the system tend to improve their social rank, the density $f(w,t)$ continuously changes in time by elementary interactions. It is clear that these interactions are more general than the trading activity described by \fer{lin}, since the social climbing is not only achieved by trading to increase the wealth, but it involves a variety of different activities finalized to this goal.
Similarly to the problems treated in  \cite{DT,GT17,GT18}, the mechanism of social climbing in modern societies can be postulated to depend on some universal features that can be summarized by saying that agents likely tend to increase the value $w$ of their ranking by interactions, while manifest a certain resistance to decrease it. 

To obtain a computable and acceptable (from the sociological point of view) expression of the elementary variation of the social ranking of agents, we identify the mean values which characterize, at least in a very stylized way, the multi-agent society. A first value, denoted by $\bar w$, defines the upper limit of the low \emph{social} ranking. This value identifies the limit below which agents do not expect to be able to climb the social ladder. The second value, denoted by  $\bar w_L$, with $\bar w_L > \bar w$, denotes the value that it is considered as the level of a satisfactory well-being by a large part of the population. Note that both these values are in general differently perceived by agents. However, since  the variations of these values from agent to agent can be assumed small, we can consider them as mean values universally perceived by agents. 

The elementary interaction is consequently modeled to describe the behavior of agents in terms of these values, and it will express the natural tendency of individuals to reach (at least) the value $\bar w_L$. 

Similarly to the case of football activity introduced in Section \ref{football}, and to the case of wealth distribution treated in the previous Section, there is a strong asymmetry in the realization of this target. With respect to individuals which enjoy a high level of social ranking, this asymmetry expresses the objective difficulty of individuals to increase the value  $w$ of the rank to reach the desired value $\bar w$ from below. 

Proceeding as in Section \ref{wealth} we will model the elementary interaction by resorting both to value functions and to uncertainties. Hence we write the elementary variation of social rank in the form
 \be\label{true-coll}
 w_* = w  - \Psi^\e\left(\frac w {\bar w_L}\right) w + \eta_\e w.
 \ee
In any interaction the value $w$ of the social rank can be modified for two different reasons, both quantified by dimensionless coefficients acting on the actual social rank variable $w$. The first one is  the (value) function $\Psi^\e(w/\bar w_L)$, which can assume assume both positive and negative values, that characterizes the asymmetric predictable behavior of agents. The second coefficients characterized a certain amount of  unpredictability always present in human activities. The uncertainty is contained in the random variable $\eta$, and , according to the wealth interactions \fer{lin}, it is designed to be  negligible in the mean, and in any case are not so significant to produce a sensible variation of the value $w$.  Hence, the social rank of individuals can be both increasing and decreasing by interactions, and the mean intensity of this variation is fully determined by the function $\Psi^\e$. 
Last, the positive parameter $\e \ll1 $ quantifies the intensity of a single interaction.

On the basis of the discussion of Section \ref{wealth}, and resorting to the class of value functions considered in  \cite{DT}, we will describe the microscopic variation of $z=1/w$, in absence of uncertainty, in the form
 \be\label{coll}
 z_* = z  - \Phi^\e\left(\frac z{\bar z_L}\right) z.
 \ee
 In  \cite{DT}, to suitably model the alcohol consumption,  the value function was assumed in the form
\fer{vvd}. Similarly to \fer{val2} the function \fer{vvd} satisfies conditions \fer{ccd} and \fer{cce} originally postulated by Kahneman and Twersky in  \cite{KT}, including the property to be positive and concave above the reference value ($s >1 $), while negative and convex below ($s <1$).\cite{DT} This function is bounded from above and below, and satisfies the bounds
  \be\label{lbb}
- \mu \,\frac{1 -e^{-\e}}{1 + e^{-\e}} \le \Phi_1^\e(s) \le 1.
 \ee
This property implies that, for small values of the parameter $\e$, the maximal amount of increase of $z$ is of the order of $\e$.  
The function  \fer{vvd} belongs to the class of value functions 
 \be\label{vd}
 \Phi_\delta^\e(s) = \mu \frac{e^{\e(s^\delta -1)/\delta}-1}{e^{\e(s^\delta -1)/\delta}+1 } , \quad  s \ge 0,
 \ee
with $0 < \delta \le1$, that interpolate between $\Phi_0^\e(s)$ and $\Phi_1^\e(s)$, where
\be\label{v2}
 \Phi_0^\e(s) = \mu \frac{s^\e -1}{s^\e +1} , \quad  s \ge 0,
 \ee
The class of value functions \fer{vd} describe the situation in which it is easier to increase the value of $z$ than to decrease it, in the whole range of possible values of the variable $z$. Then, the Fokker--Planck equations generated  in the grazing limit of interactions based on value functions like \fer{vd} correctly possess a steady state with thin tails.

\begin{figure}\centering
	{\includegraphics[width=5.5cm]{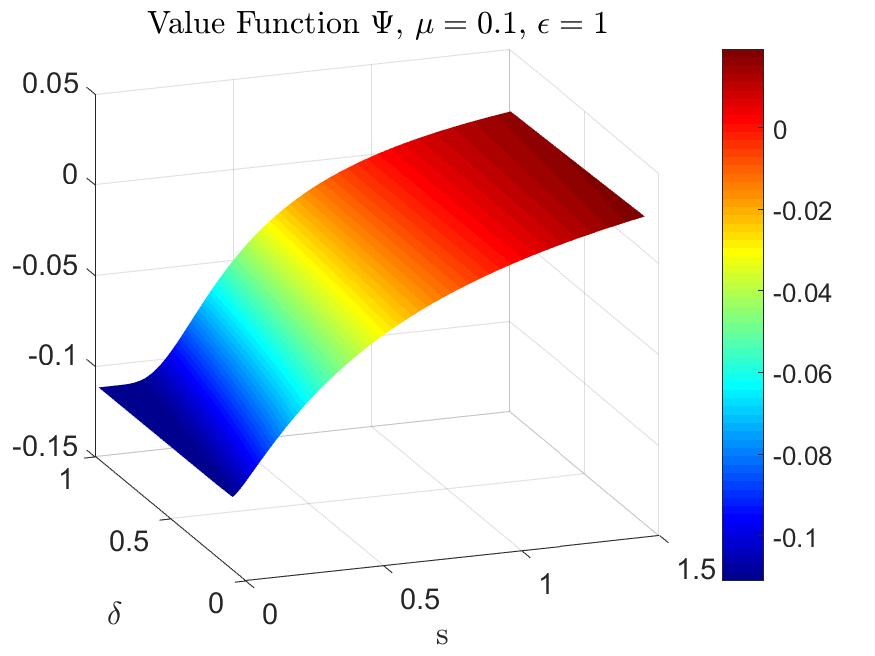}}
	\hspace{+0.35cm}
	{\includegraphics[width=5.5cm]{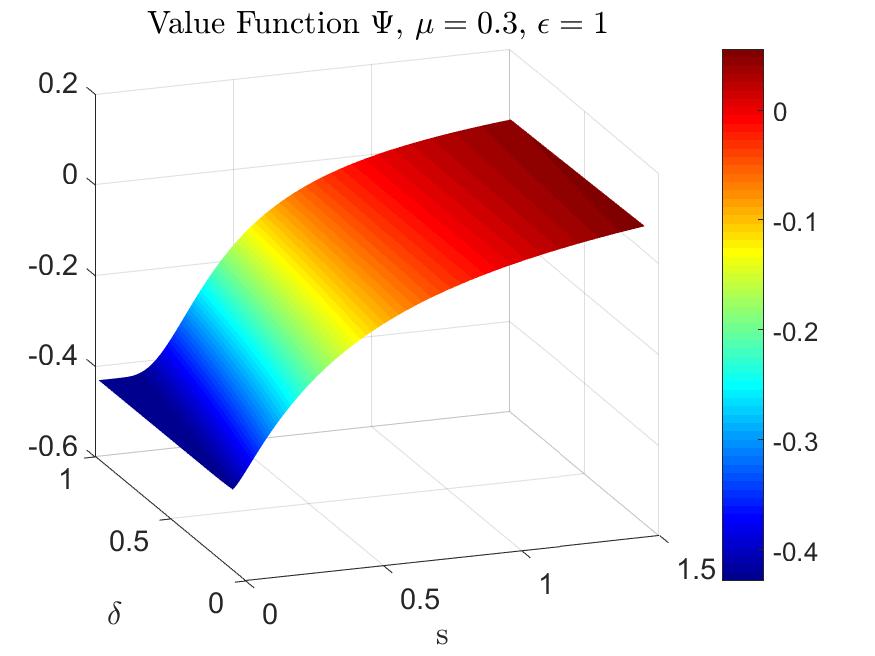}}\\
	\vspace{+0.45cm}
	{\includegraphics[width=5.5cm]{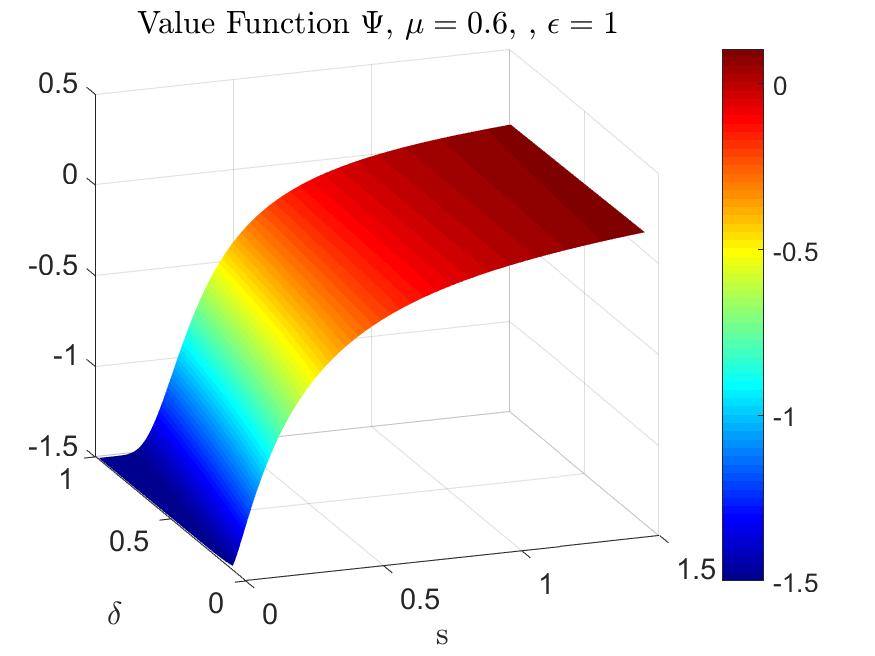}}
\hspace{+0.35cm}
{\includegraphics[width=5.5cm]{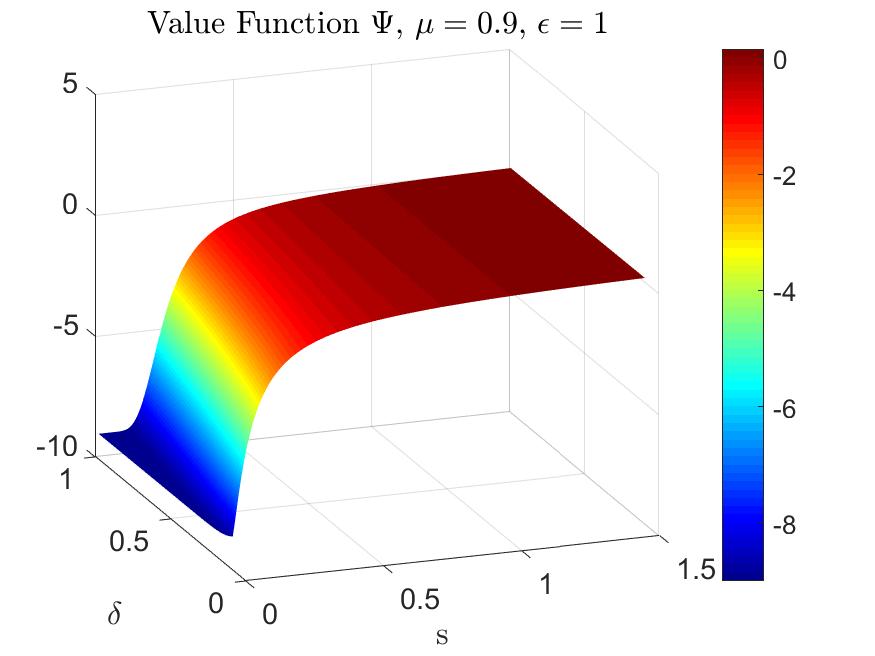}}\\
	\caption{Value function as a function of $\delta$ and $\mu$.}\label{fig1}
\end{figure}

As discussed in Section \ref{wealth}, a correct value function for the problem of the social climbing is then obtained from \fer{coll} by writing it with respect to $w = 1/z$. If the value function $\Phi^\e$ in \fer{coll} is given by \fer{vd}, one obtains that the elementary variation of the social rank value $w$ is given by \fer{true-coll}, where, if $s = w/\bar w_L$
 \be\label{vff}
 \Psi^\e(s) = \Psi_\delta^\e(s) = - \mu \frac{e^{\e(s^{-\delta} -1)/\delta}-1}{(1-\mu)e^{\e(s^{-\delta} -1)/\delta}+1+\mu } , \quad  s \ge 0.
 \ee
The value function  $\Psi_\delta^\e$ is bounded from below and above, and satisfies the bounds
 \be\label{bbb}
- \frac\mu{1-\mu}  \le \Psi_\delta^\e(s) \le \mu \frac{1 - e^{-\e/\delta}}{(1-\mu)e^{-\e/\delta}+1+\mu }.
 \ee
Starting from $s=0$, for any fixed value of the positive parameters $\e$ and $\delta$, the function $\Psi_\delta^\e(s)$ is convex in a small interval contained in the interval $(0,1)$, with an inflection point in $\bar s <1$, then concave.  A picture of the value function for different values of the parameter is shown in Figure \ref{fig1}.

\begin{figure}\centering
	{\includegraphics[width=5.5cm]{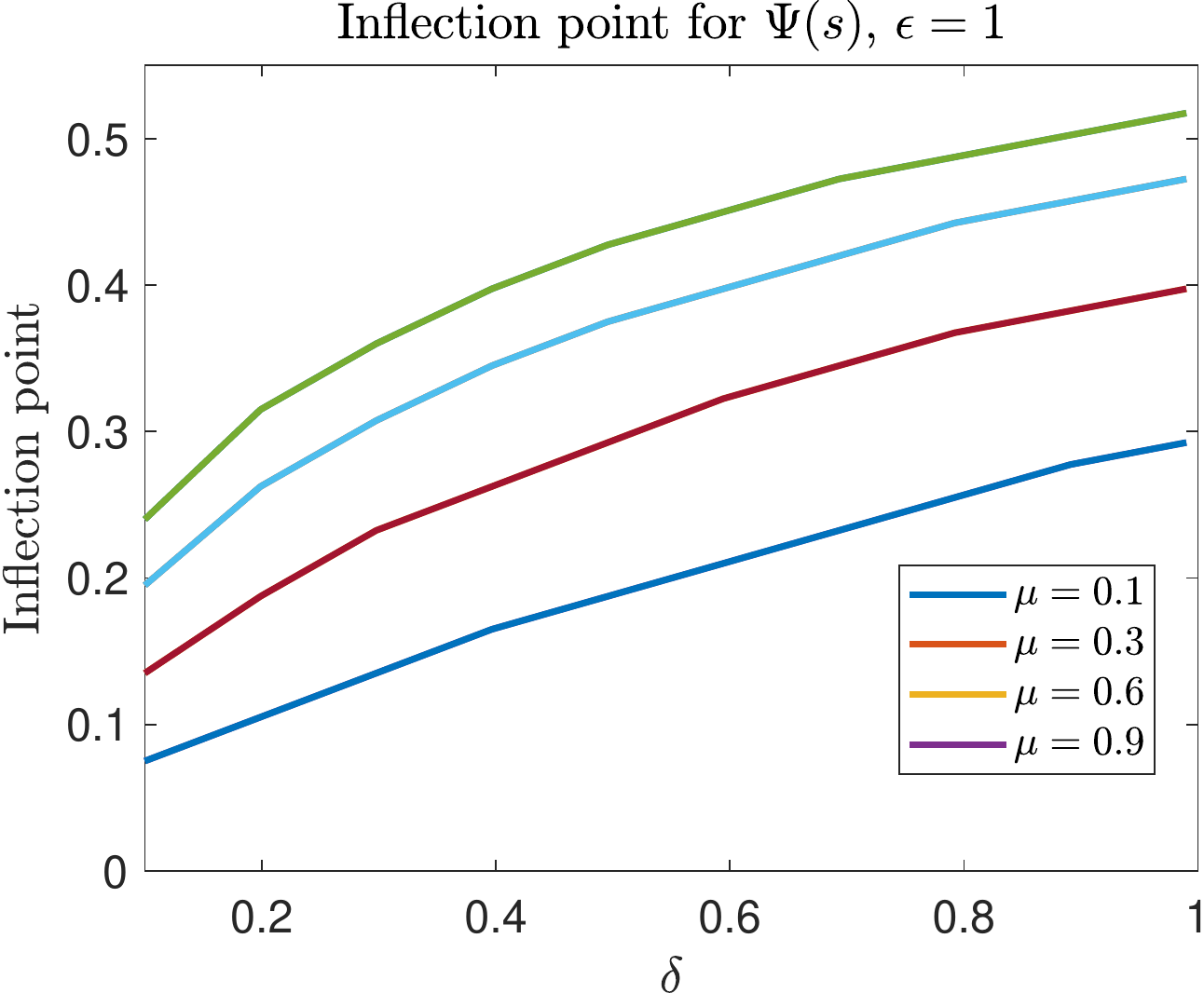}}
	\caption{Inflection points for the value function as a function of $\delta$ and $\mu$.}\label{fig2}
\end{figure}

We can now relate the inflection point $\bar s <1 $ and the reference point $s =1$ to the previously mentioned mean values characterizing the social climbing. Clearly, the value $w = \bar w_L$, namely the perceived level of a satisfactory well-being corresponds to the reference point $s=1$.  Then, the upper limit of the low social ranking $\bar w$ characterizes the inflection point $\bar s <1$, in view of the relationship
 \be\label{www}
 \bar s = \frac{\bar w}{\bar w_L} < 1.
 \ee
Indeed, the graph of the function $\Psi(s)$ can be split in the three regions $(0,\bar s)$, $(\bar s, 1)$ and $(1, +\infty)$, such that the graph is steeper in the middle region than in the other two, thus characterizing the middle region (the region of the middle social ranks) as the region in which  a higher value of the mobility is present, and it is easier  to improve the rank. On the contrary, the mobility is lower for values of the rank below $\bar w$ and above $\bar w_L$. In Figure \ref{fig2}, we show the location of the inflection points as a function of $\delta$ for different values of $\mu$. 
This behavior of the function $\Psi^\e$ is clearly in agreement with the original believe of Pareto.\cite{Par2}  

It is interesting to remark that, in the limit $\delta \to 0$  the value function \fer{vff} becomes
 \be\label{logn}
  \Psi_0^\e(s) = \mu \frac{s^\e -1}{(1+\nu)s^\e +1-\mu} , \quad  s \ge 0,
 \ee
namely a value function of the same type of \fer{v2}. Note however that in this case the limit value function \fer{logn}, at difference with the value functions \fer{vff}, is concave, and the inflection point $\bar w$ is lost.  These value functions were originally considered in  \cite{GT17,GT18}, where it was shown that they characterize the lognormal distribution.  Hence, the lognormal distribution can be defined as a border case that separates fat tailed distributions from thin tailed ones.

\subsection{The kinetic model}
Once  the \emph{interaction}  \fer{true-coll} has been  modeled, for any choice of the value function $\Psi^\e$ in the class \fer{vff} the study of the
time-evolution of the statistical distribution of the social rank follows by
resorting to kinetic collision-like models \cite{Cer,PT13}.  For any given value of the small parameter $\e$, the variation of the  density $f_\e(w,t)$  is easily shown to obey to a linear
Boltzmann-like equation, fruitfully written
in weak form. The weak form corresponds to say that the solution $f(w,t)$
satisfies, for all smooth functions $\varphi(w)$ (the observable quantities)
 \begin{equation}
  \label{kin-w}
 \frac{d}{dt}\int_{\R_+}\varphi(w)\,f_\e(w,t)\,dw  = 
  \Big \langle \int_{\R_+} \chi\left(\frac w{u}\right)\, \bigl( \varphi(w_*)-\varphi(w) \bigr) f_\e(w,t)
\,dw \Big \rangle.
 \end{equation}
 Here expectation $\langle \cdot \rangle$ takes into account the presence of the random parameter $\eta_\e$ in \fer{true-coll}. In \fer{kin-w} the positive dimensionless function $\chi(w/u)$ measures the interaction frequency of the interactions with social rank $w$ with respect to the some unit of measure $u$.

The right-hand side of equation \fer{kin-w} quantifies the variation of the observable quantity $\varphi$  of the agents that modify their value from $w$ to $w_* $  according to the elementary interaction \fer{true-coll}. 

The importance of a variable collision frequency has been outlined in  \cite{FPTT1} for the kinetic description of wealth distribution.  
There, starting from a careful analysis of the microscopic economic transactions of the kinetic model, expressed by \fer{lin}, allowed to conclude that the choice of a constant collision kernel included as possible also interactions which human agents would exclude \emph{a priori}. This was evident for example in the case of interactions in which an agent that trades with a certain amount of wealth, does not receive (excluding the risk) some wealth back from the market. In strong analogy with the rarefied gas dynamics \cite{Bob},  where the analysis of the Boltzmann equation for Maxwell pseudo-molecules leads to the possibility to make use of the Fourier transformed version,  this makes clear that, in the socio-economic modeling, the main advantages of the Maxwellian assumption are linked to the possibility to obtain analytical results. 

In agreement with  \cite{FPTT1}, and treasuring the remarks about the frequency of interactions made in Section \ref{football}, we  express the mathematical form of the kernel $\chi(\cdot)$
by assuming that, for a given value $w$ of the social rank,  the frequency of interactions trying to improve the rank is  directly proportional to  the rank itself.    This choice leads to consider collision kernels in the form
 \be\label{Ker}
 \chi\left(\frac w{u}\right) = \left( \frac w{u}\right)^{\beta} \cdot \alpha ,
  \ee
for some constants $\alpha >0 $ and $\beta >0$.  This kernel assigns a low frequency to interactions in which individuals have a low rank, and assigns a high frequency to interactions in which the value of the rank is greater. This assumption translates in a simple mathematical form that the motivations to climb the social ladder are stronger in individuals belonging to the middle and upper classes, which clearly see, at difference with individuals belonging to the low class,  the possibility to succeed.

The values of the constants $\alpha$ and $\beta$ need to be suitably chosen to guarantee at best that the characteristics of both the value function and the collective phenomenon are maintained even for small values of the parameter $\e$, and do not disappear in the limit $\e \to 0$ .

To simplify computations,  and to retain the essentials of the reasoning, let us set the unit of measure $u = \bar w_L$. Consequently
 \be\label{Ker1}
 \chi\left(\frac w{u}\right) = \alpha \cdot \left(\frac{w}{\bar w_L}\right)^{\beta} = \alpha \, s^\beta,
  \ee
For a given value function $\Psi_\delta^\e(s)$, the individual rate of growth is given by
 \be\label{gro}
 \frac{\partial \Psi_\delta^\e(s)}{\partial s} = 2 \mu  \frac \e {\left[ (1-\mu)e^{y/2} + (1+\mu)e^{-y/2}\right]^2} s^{-(1+\delta)},
  \ee  
 where $y$ is defined by
  \[
  y= \e\frac{s^{-\delta}- 1}\delta.
  \]
Now, consider that, since $\e <1$, for any given $s \ge 0$,
  \[
y  \ge  - \frac\e\delta \ge - \frac 2\delta, 
  \]
 and the function
 \[
 z(y) = \left[ (1-\mu)e^{y/2} + (1+\mu)e^{-y/2}\right]^2
 \]  
 has a maximum in the point
 \[
 \bar y = \log \frac {1+\mu}{1-\mu},
 \]
 where
 \[
 z(\bar y) = 4(1-\mu)^2.
 \]
 Therefore we easily conclude with the bounds
 \begin{equations}\label{bb3}
c_\delta &=  \frac 1{\left[ (1+\mu)e^{1/\delta} + (1-\mu)e^{-1/\delta}\right]^2} \le \\
& \frac 1{\left[ (1-\mu)e^{y/2} + (1+\mu)e^{-y/2}\right]^2} \le C_\delta = \frac 1{4(1-\mu^2)}.
 \end{equations}
 This implies that, for a given $s >0$, the individual growth is vanishing as $\e \to 0$.  To maintain a collective growth different from zero as $\e \to 0$, it is enough to fix
  \be\label{cho}
 \alpha = \frac 1{\tau\e}
 \ee
This corresponds to increase the frequency at the order $1/\e$, and to introduce at the same time a relaxation parameter $\tau$. 
 Once the choice of $\alpha$ has been justified to retain a collective memory of the growth of the value function, one has to face with the collective variation of the social rank, now given by
 \[
\frac 1{\tau\e}\Psi_\delta^\e(s) s^\beta.
 \]
Let us consider first the case $s \le 1$, namely to the increasing of the social rank. Since $\Psi_\delta^0\left(s\right)= 0$, Lagrange theorem   implies
  \[
  \frac 1\e \Psi_\delta^\e\left(s\right) =  \frac{\Psi_\delta^\e\left(s\right) - \Psi_\delta^0\left(s\right)}\e = 
  \left. \frac{\partial \Psi_\delta^\e\left(s\right)}{\partial\e}\right|_{\e = \bar \e}, \quad 0 \le \bar\e \le \e.
  \]
  Then
  \be\label{de1}
  - \frac{\partial \Psi_\delta^\e\left(s\right)}{\partial\e} = 2\mu \frac 1{\left[ (1-\mu)e^{y/2} + (1+\mu)e^{-y/2}\right]^2}\frac 1\delta\left(s^{-\delta}-1\right) >0.
  \ee
Using the upper bound in \fer{bb3} we can conclude that, when $s <1$, with the choice $\beta =\delta$ the collective variation of the social rank is uniformly bounded with respect to $\e$, and satisfies the bound
 \[
\left| \frac 1\e \Psi_\delta^\e\left(s\right) \right| s^\delta  \le \frac\mu{2\delta(1-\mu^2)}(1-s^\delta).
 \]
Note that, if we choose  $\beta < \delta$, the variation of the collective growth of social rank is extremely high for values of $s$ close to zero, namely in the part of population that do not possess many possibilities to succeed. Hence, this choice is something that we have to exclude a priori. Also, values of $\beta >\delta$ would imply that for values of $s$ close to zero, the collective growth is extremely small, thus practically excluding the possibility to increase the social rank starting from very low levels. Hence, the choice $\beta = \delta$ represents a good compromise for most societies.

With these assumptions, the weak form of the Boltzmann-type equation \fer{kin-w} takes the form
 \begin{equation}
  \label{kin}
 \frac{d}{dt}\int_{\R_+}\varphi(w)\,f_\e(w,t)\,dx  = \frac 1{\e\tau}\,\,
  \Big \langle \int_{\R_+}\left(\frac wu\right)^{\delta} \bigl( \varphi(w_*)-\varphi(w) \bigr) f_\e(w,t)
\,dw \Big \rangle.
 \end{equation}
 Note that, in consequence of the choice made on the interaction kernel $\chi$, the evolution of the density $f_\e(x,t)$ is tuned by the parameter $\e$, which characterizes both the intensity of interactions and the interaction frequency.

Due to the presence in the microscopic interaction  \fer{true-coll} of the nonlinear value function \fer{vff}, it is immediate to show that the only conserved quantity of equation \fer{kin} is obtained in correspondence to $\varphi = 1$. This conservation law implies that the solution to \fer{kin} remains a probability density for all subsequent times $t >0$.  The evolution of higher moments is not known analytically, and it is quite difficult to obtain explicit bounds, which could guarantee the precise value of the expected tail of the stationary solution.

\section{Quasi-invariant  limit and the Fokker-Planck equation}\label{quasi}

\subsection{The grazing  limit}

The linear kinetic equation \fer{kin-w} describes the evolution of the density consequent to interactions of type \fer{true-coll}, and it is valid for any choice of the parameters $\delta, \mu$ and $\e$. In real situations, however, it is reasonable to assume that in most cases a single interaction  determines only an extremely small change of the value $w$. This is certainly true for the deterministic part of the interaction \fer{true-coll}, and it is true in the mean for the random part. This situation is well-known in kinetic theory of rarefied gases, where interactions of this type are called \emph{grazing} collisions \cite{PT13,Vi}. In the value functions \fer{vd} the smallness assumption requires to fix $\e \ll 1$. At the same time \cite{FPTT},  the balance of this smallness with the random part is achieved by setting 
 \be\label{scal}
  \eta_\e = \sqrt\e \eta,
 \ee
where $\eta$ is a centered random variable of variance equal to $\sigma$. In this way the variance of $\eta_\e$ is of order $\e$.  Consequently, the scaling assumption \fer{scal} allows to retain the effect of all parameters in \fer{coll} in the  limit procedure. An exhaustive discussion on these scaling assumptions can be found in  \cite{FPTT} (cf. also  \cite{GT18} for analogous computations in the case of the lognormal distribution, in which the value function is given by \fer{v2}). For these reasons, we address the interested reader  to these review papers for exhaustive details. 

In what follows we only deal with the main differences in the computations, essentially due to the presence of a non-Maxwellian interaction kernel, and to the presence of a new type of value functions. Without loss of generality, in the rest of the paper we fix in \fer{Ker} the value of the unit of measure to $u=1$. Note that a different choice will simply correspond to a scaling of time. 

Let us suppose that the solution to the kinetic equation \fer{kin} has moments bounded up to the order two.
To have a precise idea of the evolution of the observable in the scaling $\e \ll 1$,  we start by studying the evolution of the mean value 
 \[
 m_\e(t) = \int_{\R_+} w \, f_\e(w,t) \, dw.
 \]
In this simple case we have
 \be\label{mm1}
\frac{d }{dt}\, m_\e(t) = - \frac 1\tau \,\bar w_L^{1+\delta} \, \, \int_{\R_+} \frac 1\e \Psi_\delta^\e\left(w/\bar w_L\right) \,\left( \frac w{\bar w_L}\right)^{1+\delta}\, \, f_\e(w,t)\, dw,
 \ee
 where the integral has been written in terms of the quotient $w/\bar w_L$. 
 Using the bounds \fer{bb3} into \fer{mm1} we obtain
 \begin{equations} \label{mm2}
 \frac{d }{dt}\, m_\e(t)& =  \le C_\delta \, \frac 1\tau \,\bar w_L^{1+\delta} \, \, \int_0^w\frac 1\delta\left( \left( \frac{\bar w_L}w\right)^\delta -1 \right) \left( \frac w{\bar w_L}\right)^{1+\delta}\, \, f_\e(w,t)\, dw \\
 &+ c_\delta \, \frac 1\tau \,\bar w_L^{1+\delta} \, \, \int_w^\infty \frac 1\delta\left( \left( \frac{\bar w_L}w\right)^\delta -1 \right) \left( \frac w{\bar w_L}\right)^{1+\delta}\, \, f_\e(w,t)\, dw  \\
 & \le   \, \frac 1\tau \,\bar w_L^{1+\delta} \, \,\left( C_\delta \int_{\R_+} \frac w{\bar w_L}\, \, f_\e(w,t)\, dw 
 - c_\delta \int_{\R_+} \left( \frac{\bar w_L}w\right)^{1+\delta}\, \, f_\e(w,t)\, dw \right).
  \end{equations}
Finally, Jensen's inequality implies
 \[
 \int_{\R_+} \left( \frac{\bar w_L}w\right)^{1+\delta}\, \, f_\e(w,t)\, dw \ge \left(\int_{\R_+} \frac w{\bar w_L}\, \, f_\e(w,t)\, dw \right)^{1+\delta}.
 \]
Let us set
 \[
 M_\e(t) = \frac{m_\e(t) }{\bar w_L} .
 \]
Then 
$M(t)$ satisfies the differential inequality
 \be \label{dd1}
 \frac{d }{dt}\, M_\e(t) \le  \, \frac 1\tau \,\bar w_L^{\delta} \, \,\left( C_\delta \, M_\e(t) - c_\delta \, M_\e(t)^{1+\delta} \right),
 \ee
which shows that, independently of $\e$, $M_\e(t)$ satisfies the inequality
\be\label{mm4}
M_\e(t) \le \max \left\{  M_\e(t=0), \left( \frac{C_\delta}{c_\delta}\right)^{1/\delta}\right\}.
\ee
Coupling inequality \fer{dd1} with the uniform bound \fer{mm4} shows that   the scaling \fer{scal} is such that, for any given fixed time $t >0$, the consequent variation of the mean value $m_\e(t)$ is bounded from above independently of $\e$.  
Since pointwise
 \be\label{AA}
A_{\delta,\e}\left(\frac w{\bar w_L}\right) = \frac 1\e \Psi_\delta^\e\left(w/\bar w_L\right) \to \frac\mu{2\delta} \left( 1 -\left( \frac {\bar w_L}w\right)^\delta \right),
  \ee
 we have a computable evolution of the mean even in the limit $\e \to 0$. 
As explained in  \cite{FPTT}, there is a further physical  motivation leading to the scaling of the frequency of interactions, as given by  \fer{cho}. Since for $\e \ll1$  the  interactions  produce a very small change of the rank value $w$, in the limit $\e \to 0$, a finite  variation of the mean density can be observed only if each agents in the system undergo a great number of interactions  in a fixed period of time. 
By using the bounds \fer{bb3} one can observe, at the price of an increasing number of computations, the existence of bounds for  the evolution of the second moment of $f_\e(w,t)$, which remains well-defined also in the limit $\e \to 0$ (cf. the analysis in  \cite{FPTT}).  

It is now easy to justify the passage from the kinetic model \fer{kin-w} to its continuous counterpart given by a Fokker--Planck type equation \cite{FPTT}.
Given a smooth function $\varphi(w)$, and a collision of type \fer{coll} that produces a small variation of the difference $w_*-w$,  one obtains
 \[
\langle w_* -w \rangle = - \e \,A_{\delta,\e}\left(\frac w{\bar w_L}\right)w;  \quad  \langle (w_* -w)^2\rangle =  \left(\e^2 \,  \,A_{\delta,\e}\left(\frac w{\bar w_L}\right)^2 + \e \sigma\right) w^2.
 \]
Therefore, equating powers of $\e$,  it holds
 \[
\langle \varphi(w_*) -\varphi(w) \rangle =  \e \left( - \varphi'(w)\, w\, \frac\mu{2\delta} \left( 1- \left(\frac{\bar w_L}w\right)^\delta  \right)
  + \frac \sigma 2 \, \varphi''(w)  w^2 \right) + R_\e (w),
 \]
where the remainder term $R_\e(w)$, for a suitable $0\le \theta \le 1$ is such that 
  \[
  \frac 1\e \, R_\e(w) \to 0
  \]
 as  $\e \to 0$. Therefore one obtains that the time variation of the (smooth) observable quantity $\varphi(w)$  satisfies
\[
\begin{aligned}
 & \frac{d}{dt}\int_{\R_+}\varphi(w) \,f_\e(w,t)\,dw  = \\
 & \frac 1\tau \int_{\R_+} w^\delta \left( - \varphi'(w)\,w \,\frac\mu{2\delta} \left( 1 -\left(\frac {\bar w_L}w\right)^\delta  \right)   + \frac \sigma 2\, \varphi''(w) w^2 \right) f_\e(w,t)\, dw \ + \frac 1\e \mathcal R_\e(t) ,
 \end{aligned}
 \]
where $\mathcal R_\e(t)$ denotes the integral remainder term
\[
\label{rem3}
\mathcal R_\e(t) = \int_{\R_+ } w^\delta R_\e(w)  f_\e(w,t)\, dw. 
\]
 Letting $\e \to 0$,  shows that in consequence of the scaling \fer{scal} the weak form of the kinetic model \fer{kin-w} is well approximated by the weak form of a linear Fokker--Planck equation (with variable coefficients)
\begin{equations}
  \label{m-13}
 & \frac{d}{dt}\int_{\R_+}\varphi(w) \,g(w,t)\,dw  = \\
  & \frac 1\tau \int_{\R_+} \int_{\R_+} \left( - \varphi'(w) \,w^{1+\delta} \,\frac\mu{2\delta} \left( 1- \left(\frac{\bar w_L}w\right)^\delta \right)  + \frac \sigma 2 \varphi''(w) w^{2+\delta} \right) g(w,t)\, dw.
   \end{equations}  
 In \fer{m-13} the density function $g(w,t)$ coincides with the limit, as $\e \to 0$, of the density $f_\e(w,t)$ \cite{FPTT}.
Provided the boundary terms produced by the integration by parts vanish,  equation \fer{m-13} coincides with the weak form of the Fokker--Planck equation
 \begin{equation}\label{FP2}
 \frac{\partial g(w,t)}{\partial t} = \frac {\tilde\sigma} 2 \frac{\partial^2 }{\partial w^2}
 \left(w^{2+\delta} g(w,t)\right )+ \frac {\tilde\mu}{2}
 \frac{\partial}{\partial w}\left( \frac 1\delta \left( 1 - \left(\frac{\bar w_L}w\right)^\delta  \right) w^{1+\delta} \, g(w,t)\right).
 \end{equation}
In \fer{FP2} we defined $\tilde\sigma = \sigma/\tau$ and $\tilde\mu= \mu/\tau$. Equation \fer{FP2} describes the evolution of the distribution density $g(w,t)$ of the social rank $w \in \R_+$ of the agent's system, in the limit of the \emph{grazing} interactions.  As often happens with Fokker-Planck type equations, the steady state density can be explicitly evaluated, and it results to be a generalized Gamma density, with parameters linked to the details of the microscopic interaction \fer{coll}.
 
\subsection{Steady states are Amoroso distributions}

Let us set $\gamma = \tilde\mu/ \tilde\sigma = \mu/\sigma$. The statio\-nary distribution of the Fokker--Planck equation \fer{FP2} is an integrable function which solves the first order differential equation 
 \be\label{sd}
 \frac{d }{dw}
 \left(w^{2+\delta} g(w)\right )+ 
 \frac\gamma\delta\left(1 - \left(\frac{\bar w_L}w\right)^\delta  \right) w^{1+\delta} \, g(w)  =0.
 \ee
We solve \fer{sd} with respect to $h(w)= w^{2+\delta} g(w)$ by separation of variables. The function $h(w)$ solves the differential equation
 \be\label{inte}
 \frac{dh(w)}{dw} + \frac\gamma\delta \left(\frac 1w - \frac{\bar w_L^\delta}{w^{1+\delta}}\right)h(w) = 0.
 \ee
If $h(w) \not= 0$, equation \fer{inte} is clearly equivalent to
 \be\label{inte-1}
\frac 1{h(w)}\frac{dh(w)}{dw} = - \frac\gamma\delta \left(\frac 1w - \frac{\bar w_L^\delta}{w^{1+\delta}}\right),
 \ee
that can be rewritten as
 \be\label{inte-2}
 \frac{d}{dw}\log h(w) = - \frac\gamma\delta \left\{\frac d{dw}\log\frac w{\bar w_L} + \frac 1\delta \frac d{dw}\left[ \left( \frac{\bar w_L}{w} \right)^\delta -1 \right]\right\}.
 \ee
In this way we find that the unique solutions to \fer{sd}  are the functions 
 \be\label{equilibrio}
g_\infty(w) =  g_\infty(\bar w_L) \left( \frac{\bar w_L}{w} \right)^{2 +\delta + \gamma/\delta} \exp\left\{ - \frac \gamma{\delta^2}\left( \left( \frac{\bar w_L}w \right)^\delta -1 \right)\right\}.
 \ee 
 If we fix the mass of the steady state \fer{equilibrio} equal to one,  the consequent probability density is a particular case of the generalized Gamma distribution, usually named Amoroso distribution \cite{Amo}, as given in \fer{equi}. Note that \fer{equilibrio} corresponds to values of $\beta = -\delta <0$ in the expression of the generalized Gamma distribution \fer{equi}. 
The remaining parameters of the equilibrium state \fer{equilibrio} are given by  
 \be\label{para}
 \alpha =  1 + \frac 1\delta + \frac\gamma{\delta^2},  \quad  \theta = \bar w_L \left( \frac\gamma{\delta^2}\right)^{1/\delta}.
 \ee
The limit $\delta \to 0$ in the Fokker--Planck equation \fer{FP2} corresponds to the drift term induced by the value function \fer{v2}. In this case, the equilibrium distribution \fer{equilibrio} takes the form of a lognormal density \cite{GT17}. It is remarkable that the lognormal density is achieved as limit case when looking at the distribution of alcohol consumption \cite{DT}, where the Fokker--Planck type equation has been obtained resorting to the value functions \fer{vd}. Hence, the lognormal density appears as the limit case of both generalized Gamma distributions characterized by positive values of the parameter $\beta$, and the present case of Amoroso distributions, given by negative values of the parameter $\beta$.  
 
The case $\delta =1$ in \fer{equilibrio} corresponds to an inverse Gamma distribution. In this case $\alpha = 2+ \gamma$, and $\theta = \gamma \bar w_L$.  The steady state of the distribution of social rank coincides with the steady state of wealth distribution discussed in Section \ref{wealth}. Hence, in the grazing limit, the inverse Gamma distribution appears as steady state of two different kinetic models,  characterized by two different value functions \fer{vd} with $\delta =1$, and, respectively \fer{val}. 

Note that for all values of $\delta >0$ the moments are expressed in terms of the parameters $\bar w_L$, $\sigma$, $\mu$ and $ \delta$ denoting respectively the target level of social rank,  the variance $\sigma$ of the random effects and the values $ \delta$ and $\mu$ characterizing  the value function $\Psi_\delta^\e$ defined in \fer{vff}.   

It is interesting to remark that, in the case of the inverse Gamma distribution, the mean value of the equilibrium density \fer{equilibrio} is always less than $\bar w_L$. In this case in fact
 \[
 M = \bar w_L \frac{\gamma}{\gamma+1}. 
 \]

\section{Numerics}\label{numerics}
  In this Section, we perform several numerical experiments with the aim of describing the behaviors of the social climbing Boltzmann model \eqref{true-coll} and to quantify the goodness of its  Fokker-Planck counterpart. In more details, we intend to show the effects of different choices of the parameter  appearing in the value function $\Psi^\epsilon_\delta$ defined by \eqref{vff}. The shape of both the distribution function in time and its final steady state are indeed modified when $\mu/\sigma$, $\epsilon$ and $\delta$ are modified. The first quantity determines the intensity of the interaction with respect to the intensity of the random effects in the climbing dynamic. The second quantity determines the rate of change of the social status of an agent due to a single interaction while the last parameter is responsible for the final shape of the social distribution in a given population.
  
 \subsection{Trend to equilibrium}\label{numericsI}
\begin{figure}\centering
	{\includegraphics[width=5.5cm]{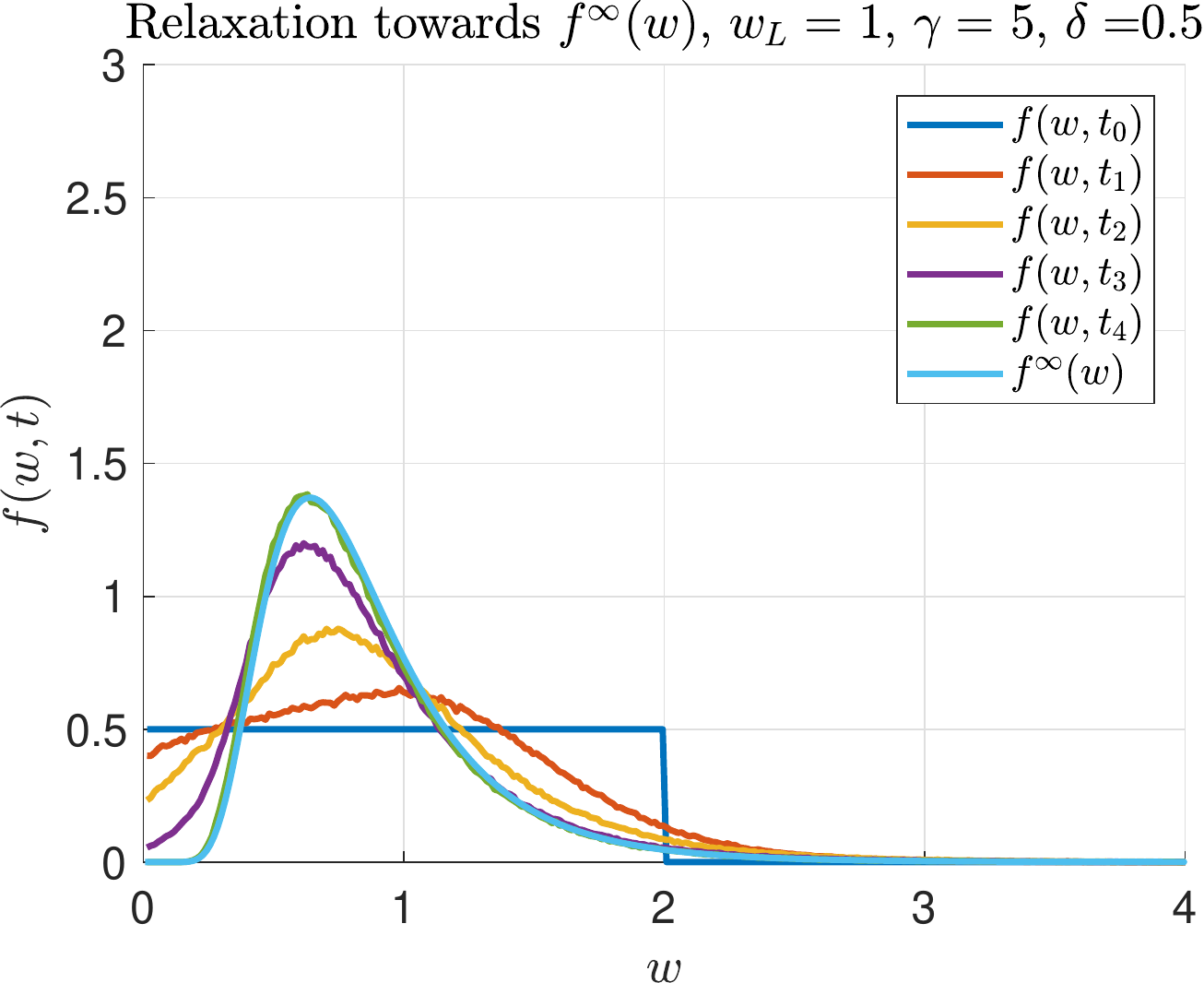}
	\includegraphics[width=5.5cm]{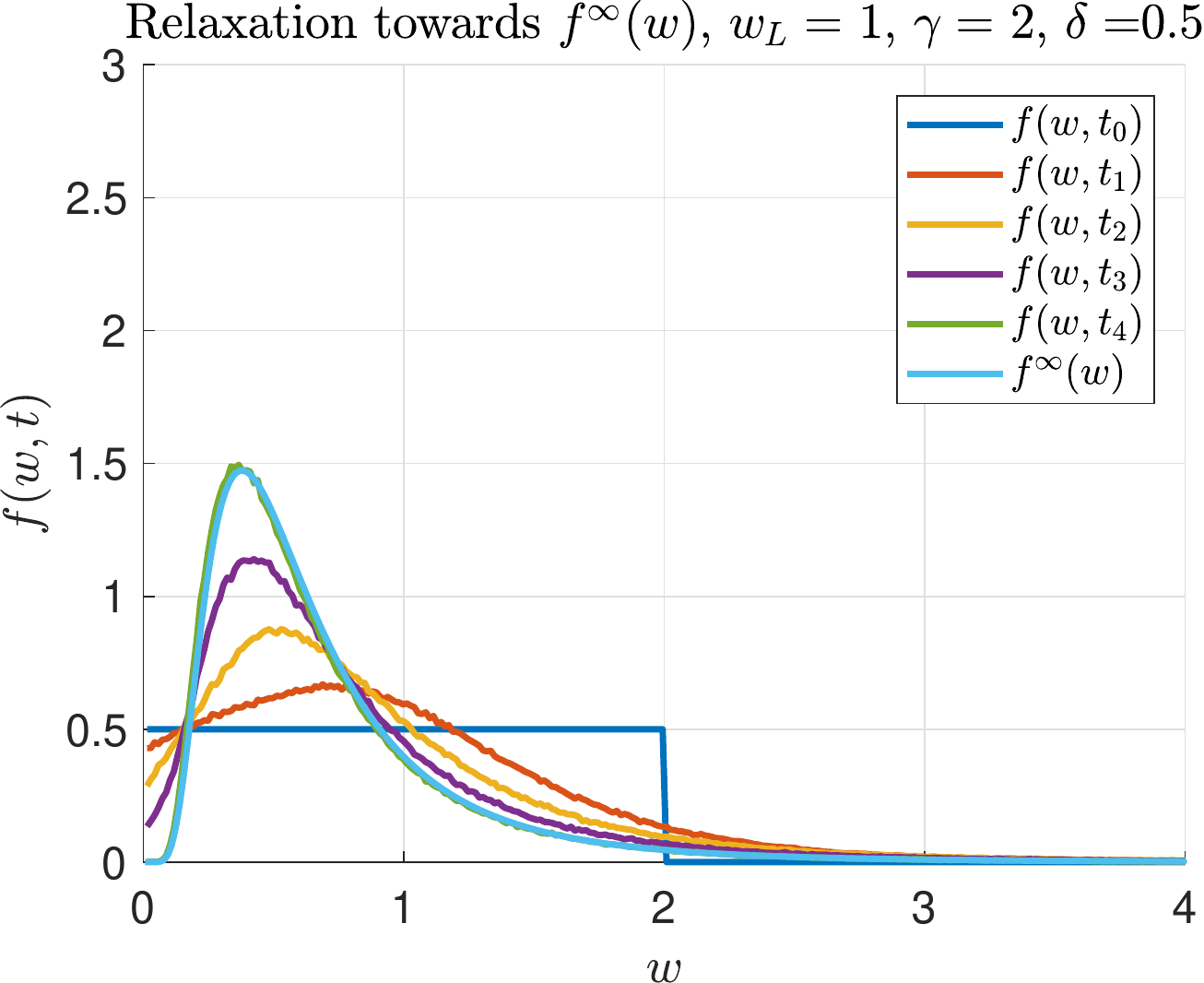}\\\vspace{1cm}
\includegraphics[width=5.5cm]{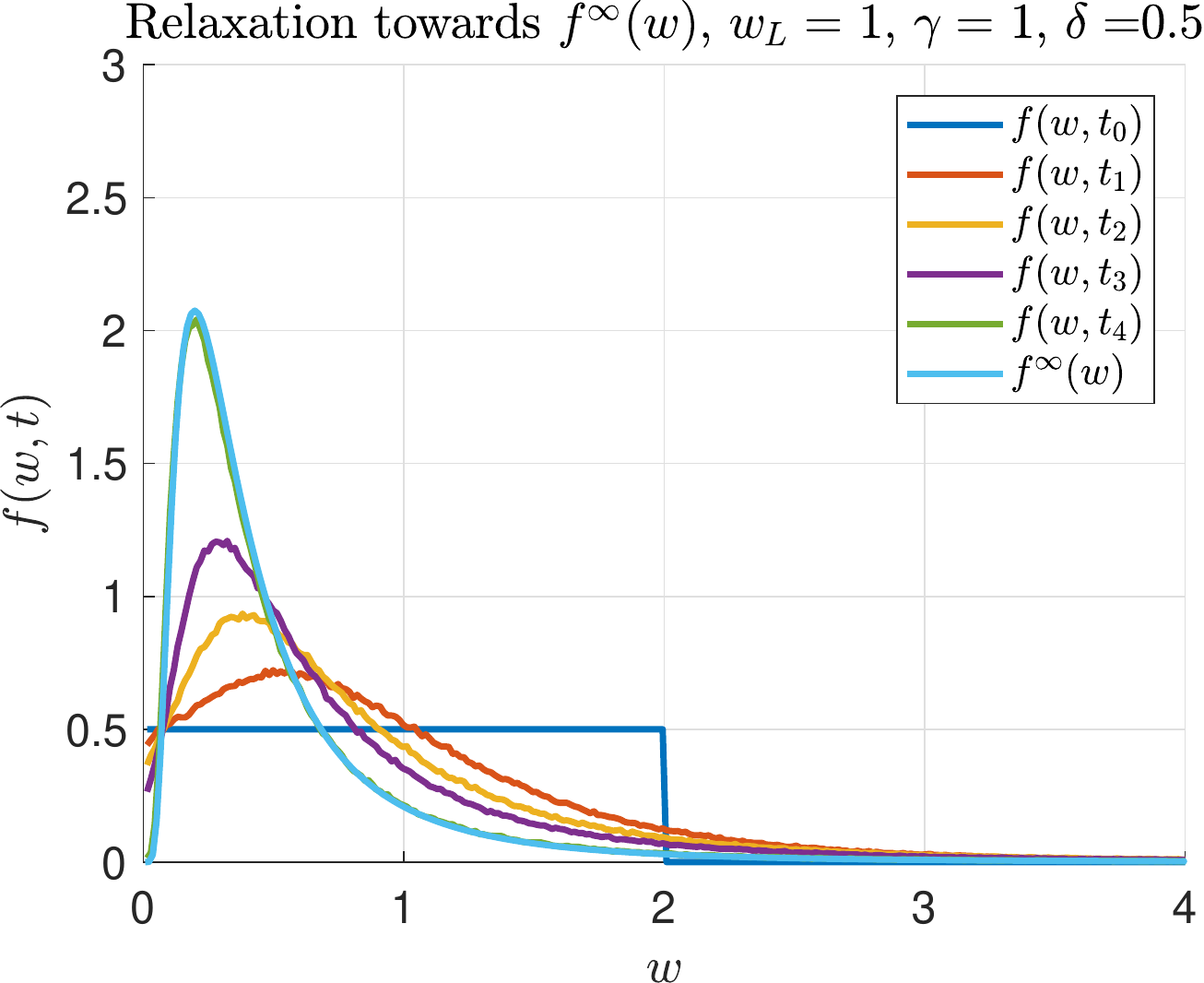}
\includegraphics[width=5.5cm]{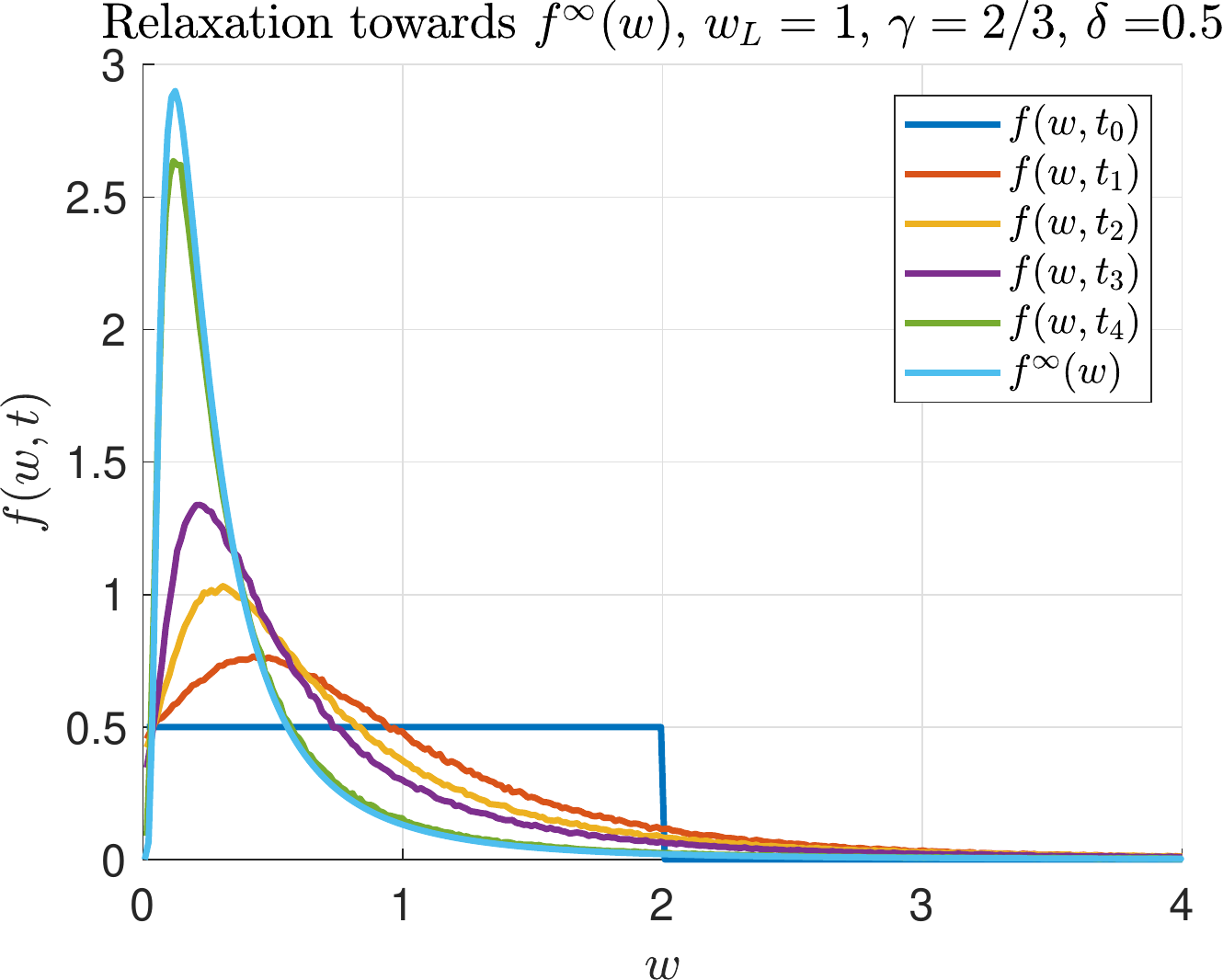}}
	\caption{Test 1. Convergence to the Fokker-Planck steady state for the Boltzmann model as a function of time with $\delta=0.5$. Top left $\gamma=5$, top right $\gamma=2$, bottom left $\gamma=1$ and bottom right $\gamma=2/3$. }\label{fig:test1}
\end{figure}

 In this first test, we analyse the rate of trend to equilibrium for the Boltzmann model \eqref{true-coll} starting from a uniform distribution $f(w,t=0)$ over the domain $[0,2]$. The convergence to the steady state Amoroso distribution for various values of the parameters $\gamma=\mu/\sigma$ are shown in Figure \ref{fig:test1}. The random effects are taken into account by sampling $\eta_\e$ from a uniform distribution with variance $\sigma$. The average level of well-being $\bar{w}_L$ is fixed equal  to $1$ while the parameter $\delta$ which characterizes the exponent of the steady state is fixed equal to $0.5$. The collision frequency is $w^\delta$, the number of samples is $10^6$ and $\varepsilon=10^{-3}$. We clearly observe that for all tested situations the Boltzmann model converges towards the corresponding Amoroso equilibrium distribution  \eqref{equilibrio}.

\subsection{Wealth and Social climbing}\label{numericsII}
\begin{figure}\centering
	{\includegraphics[width=5.5cm]{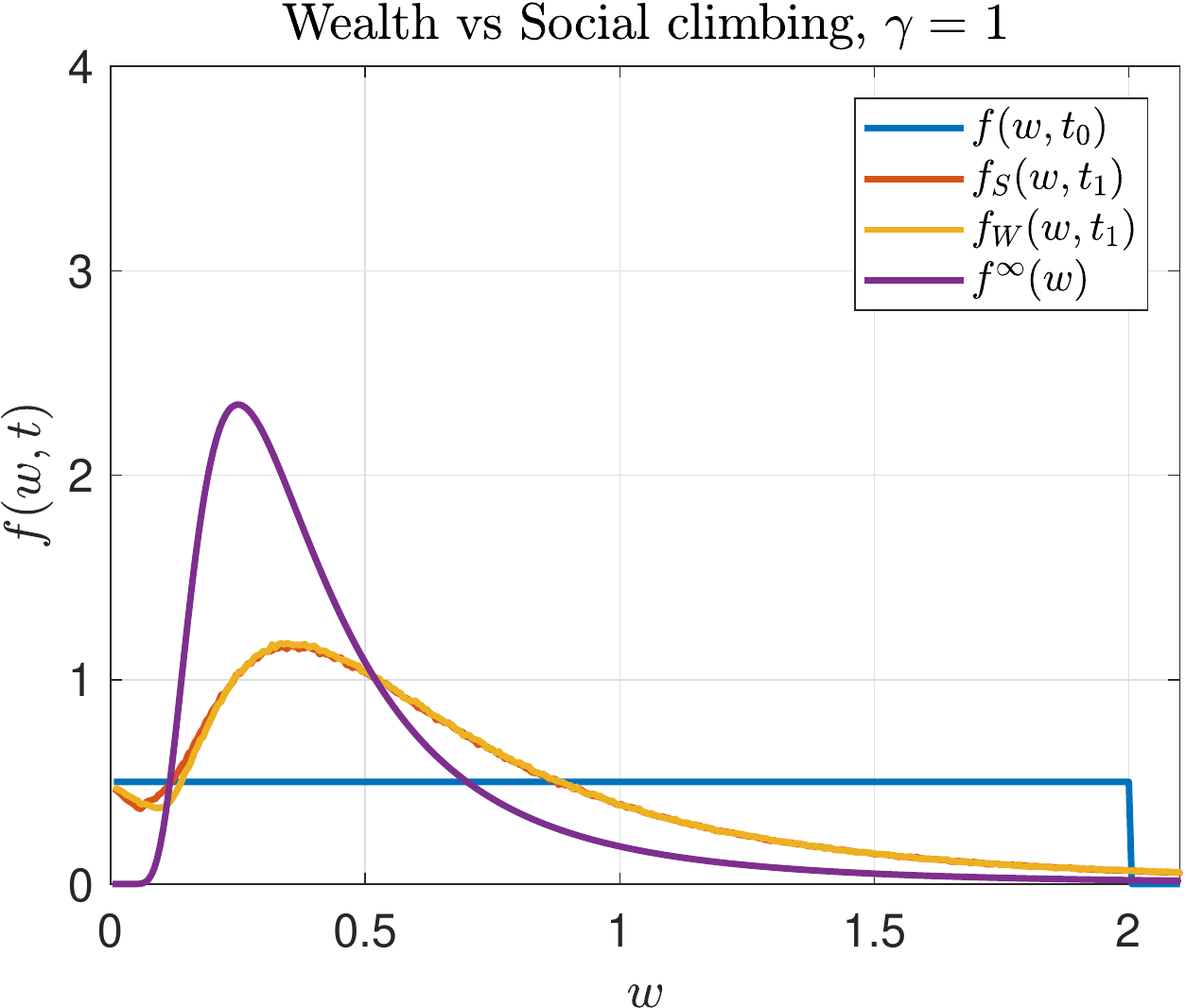}
		\includegraphics[width=5.5cm]{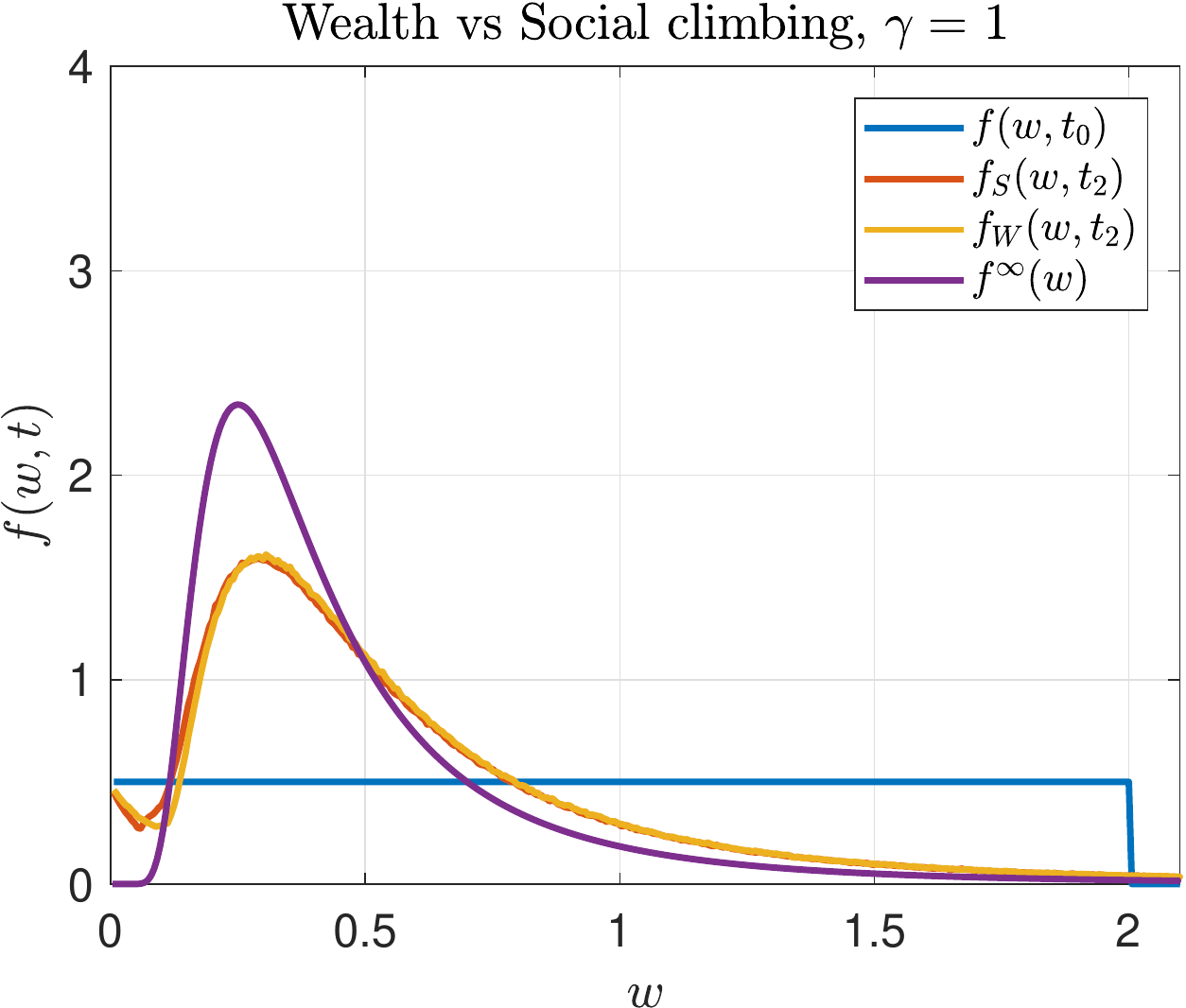}\\\vspace{1cm}
		\includegraphics[width=5.5cm]{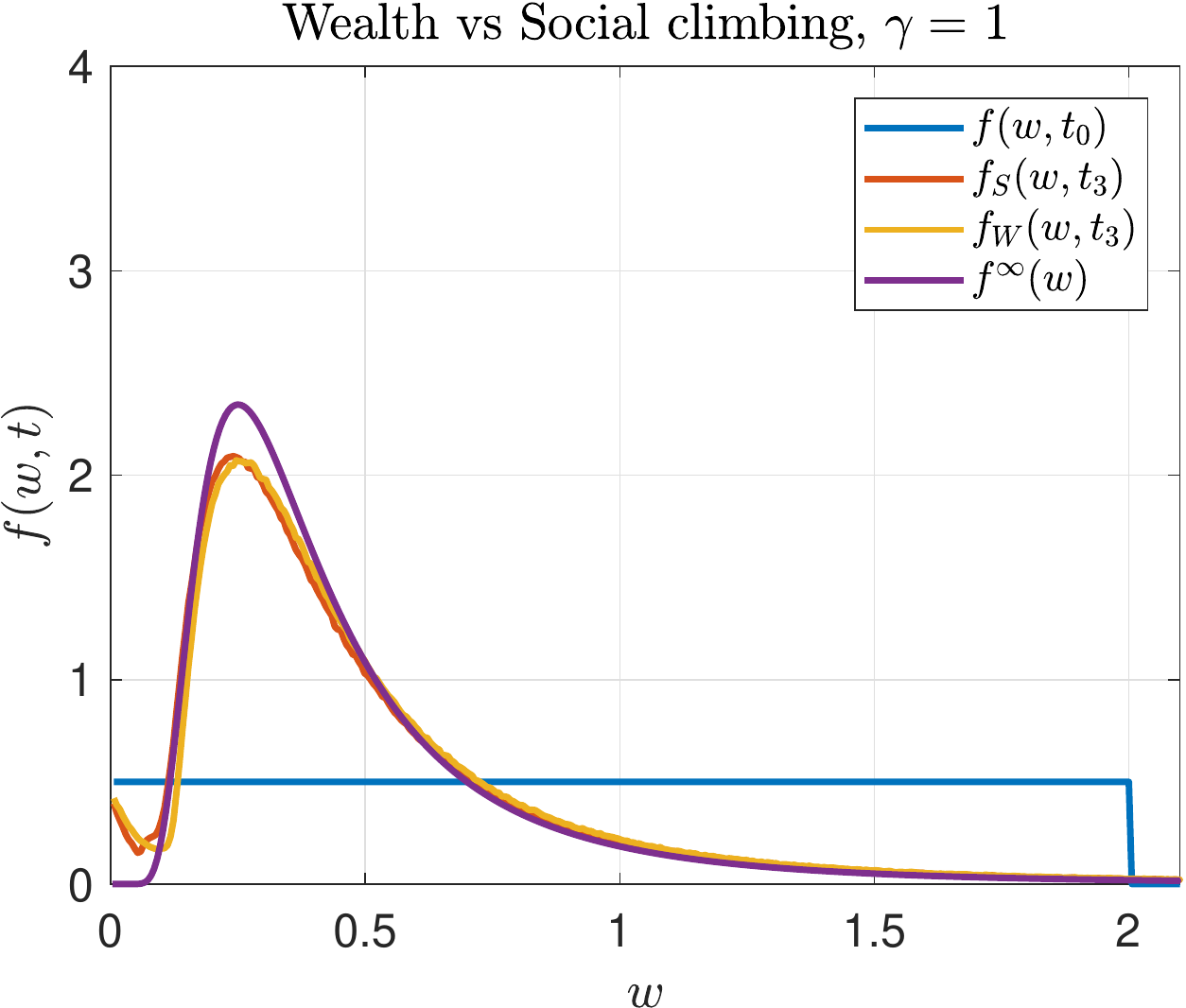}
		\includegraphics[width=5.5cm]{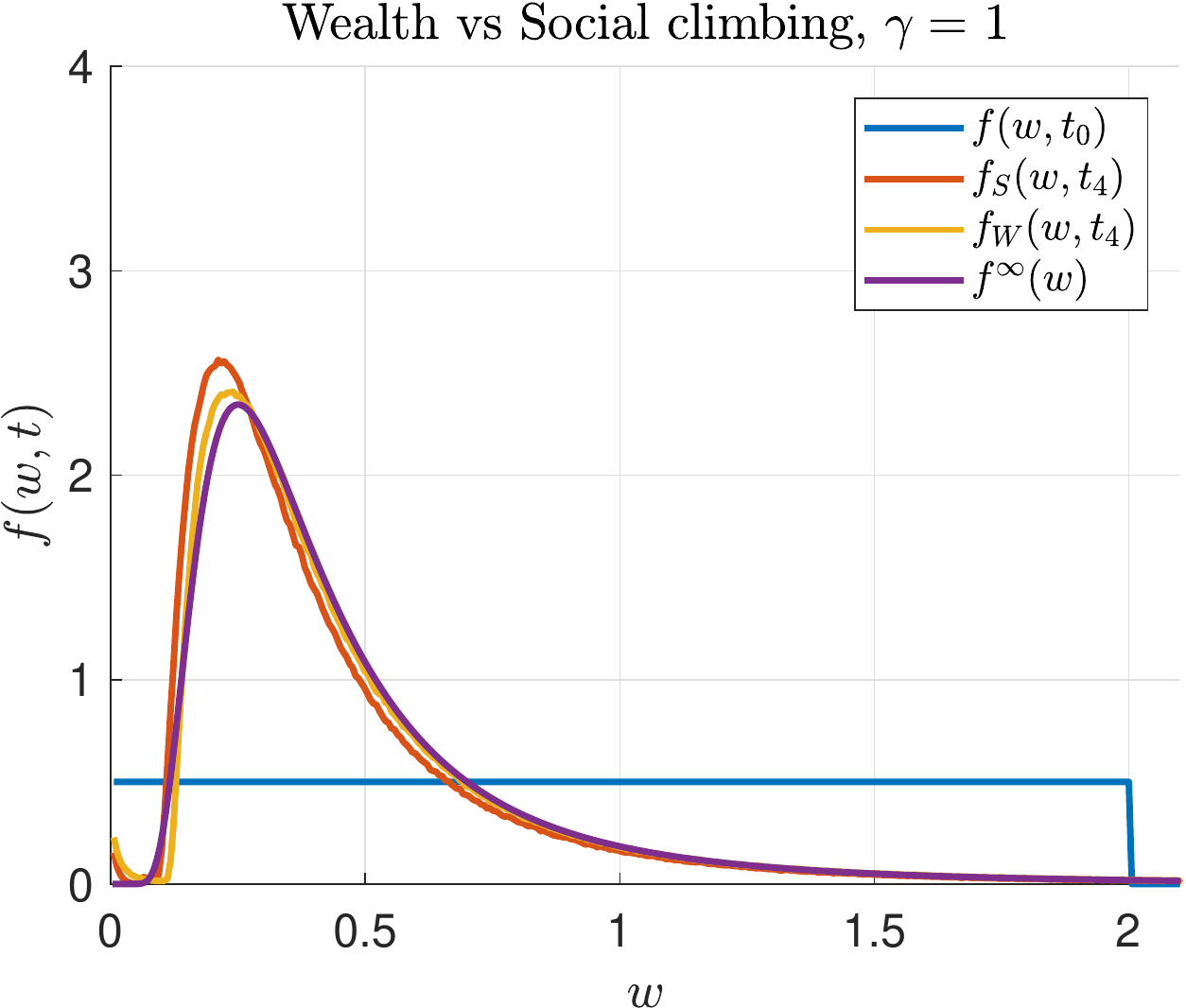}}
	\caption{Test 2. Comparison of the wealth and the social climbing models in the case $\delta=1$, $\gamma=1$ and $\epsilon=0.1$. From top left to bottom right the images show the trend to the steady state solution for the two models. The initial and the final Fokker-Planck states are also shown.}\label{fig:test21}
\end{figure}

In this part, we compare the different behaviors of  the  Boltzmann model for wealth distribution described by equation \eqref{lin2} and the Boltzmann model for social climbing \eqref{vff}. The main difference between the two models relies  in the convexity/concavity properties of the two values functions acting on the lower values of the wealth (here we identify wealth and social rank value). The first value function, describing variation of wealth, is always concave while the second, describing the social level, has an inflection point for $\bar{w}<\bar{w}_L$. This corresponds to the empirical observation that for people belonging to the lower level of the social ladder, the change of status is very difficult. For the tests reported in Figure \ref{fig:test21}, we have chosen $\delta=1$, which gives an inverse Gamma distribution as common steady state solution for the Fokker-Planck equation \eqref{FP2}. The collision frequency of the wealth model is chosen accordingly to  \cite{FPTT1} as $(vw)^\delta$ where $v$ is sampled from a uniform distribution around $\bar{w}_L$. Moreover $\gamma=1$, $\bar{w}_L=1$ while the scaling parameter $\epsilon=0.1$ is such that the models \eqref{lin2} and \eqref{vff} are far from the corresponding Fokker-Planck limits.
 
\begin{figure}\centering
	{\includegraphics[width=5.5cm]{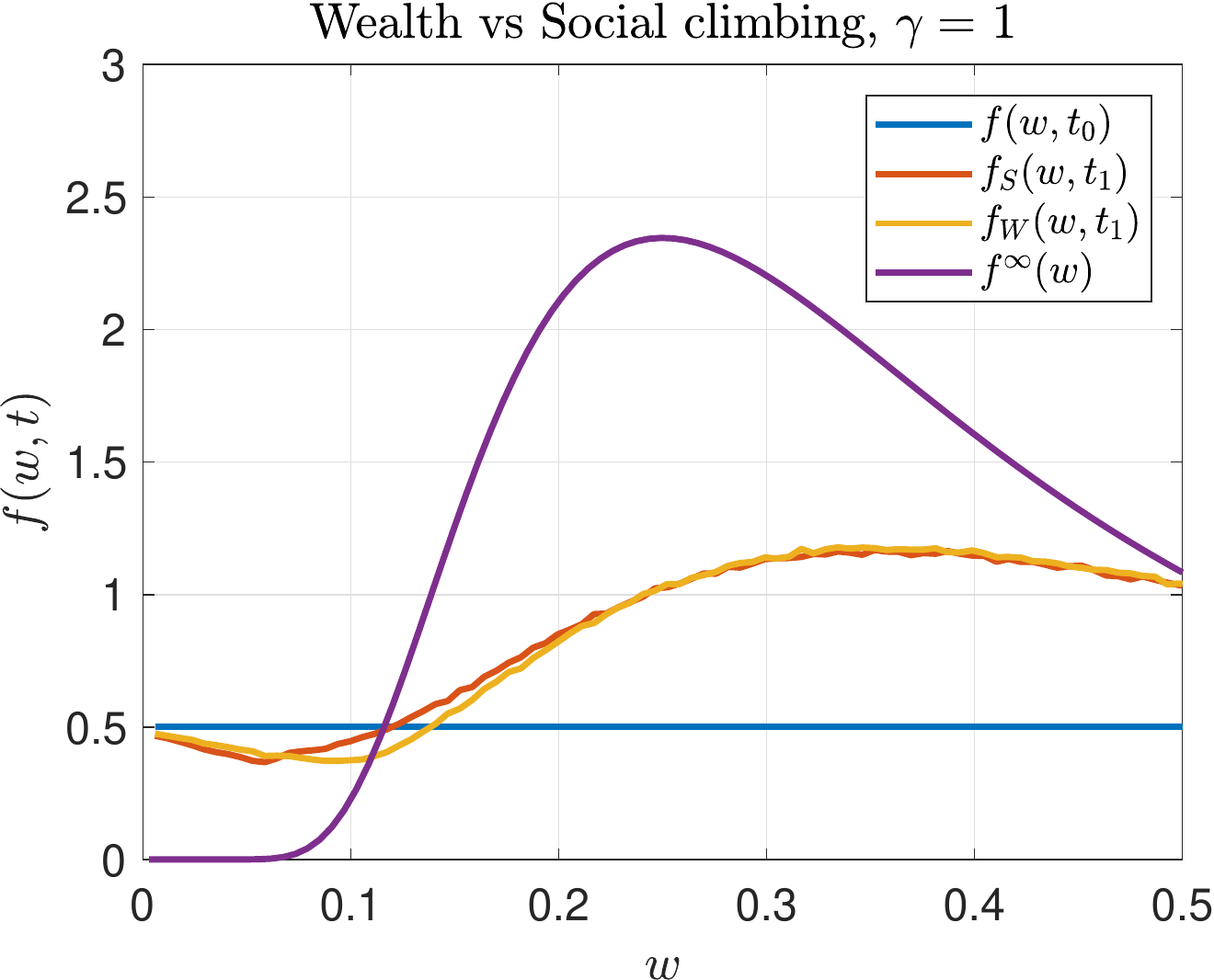}
		\includegraphics[width=5.5cm]{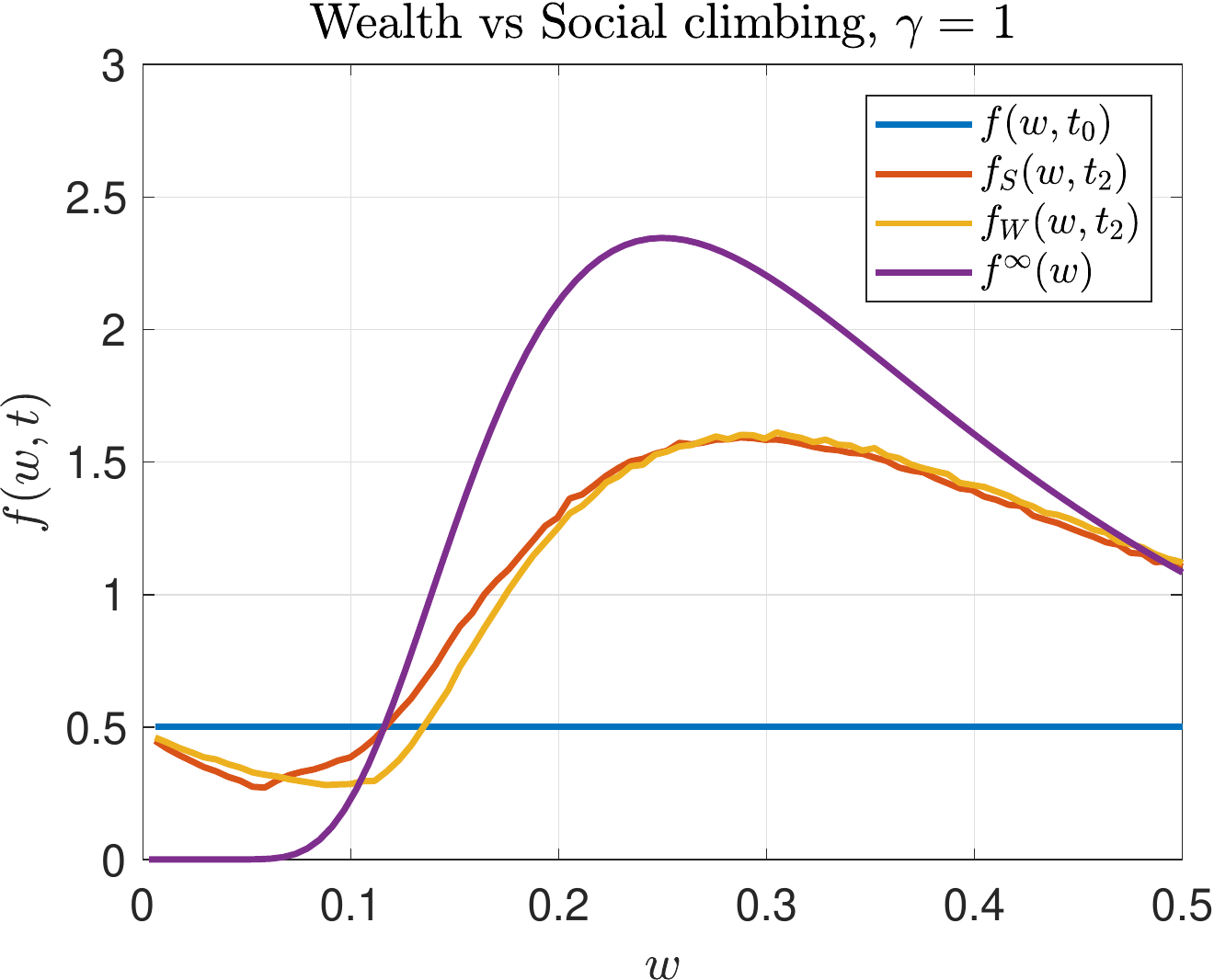}\\\vspace{1cm}
		\includegraphics[width=5.5cm]{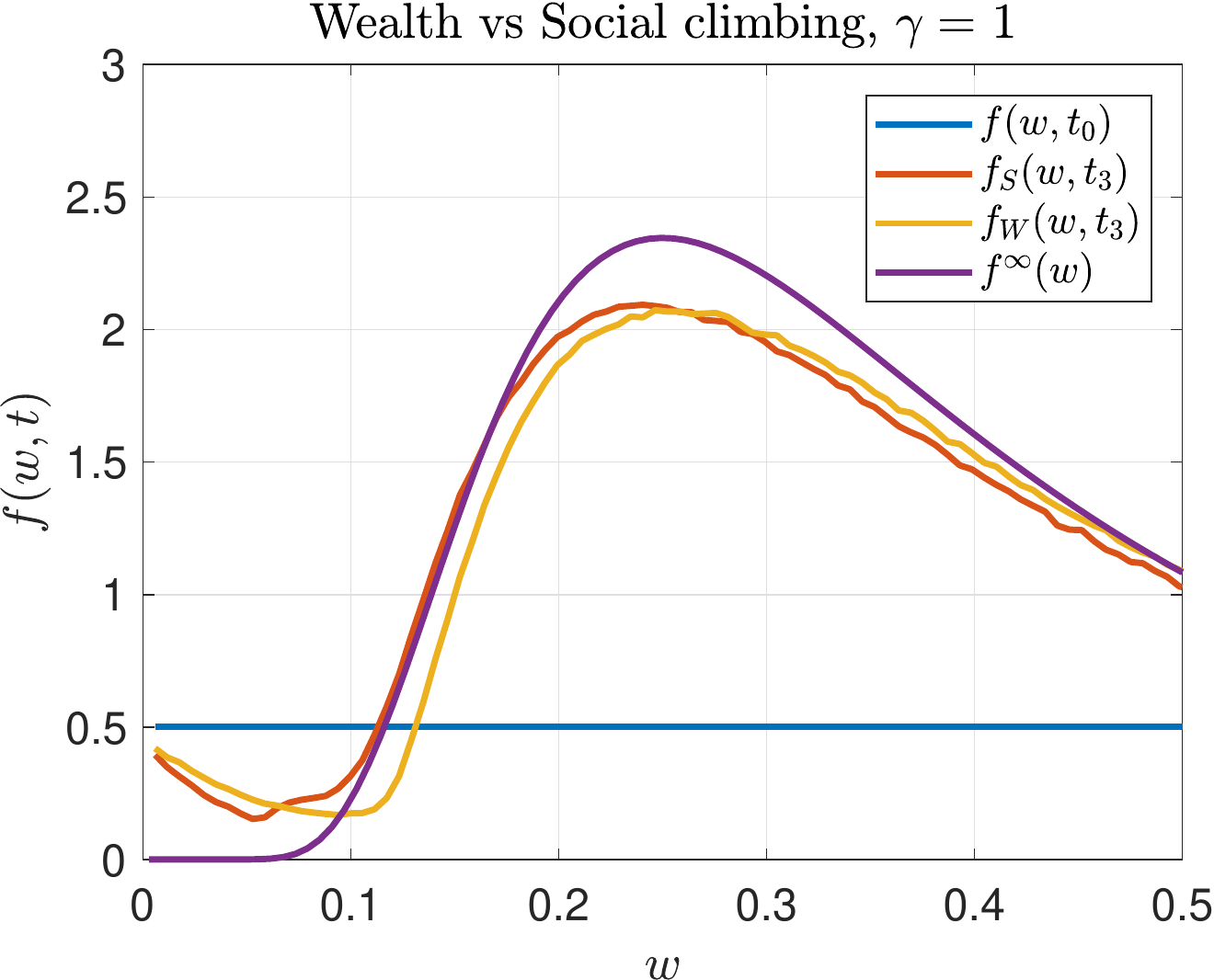}
		\includegraphics[width=5.5cm]{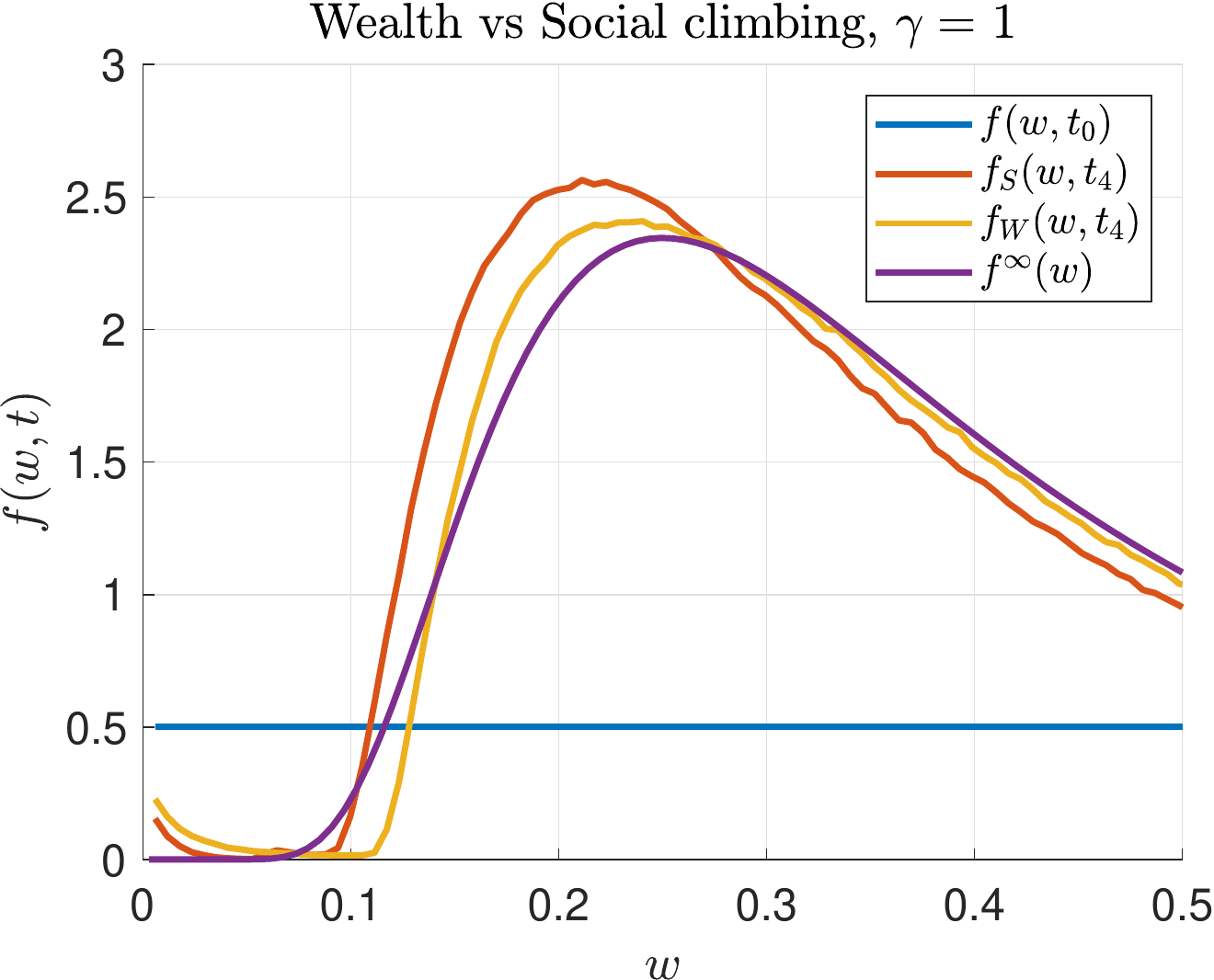}}
		\caption{Test 2. Comparison of the wealth and the social climbing models in the case $\delta=1$, $\gamma=1$ and $\epsilon=0.1$. From top left to bottom right the images show the trend to the steady state solution for the two models. The initial and the final Fokker-Planck states are also shown. Zoom around the lower wealth levels.}\label{fig:test21z}
\end{figure}

\begin{figure}\centering
	{\includegraphics[width=5.5cm]{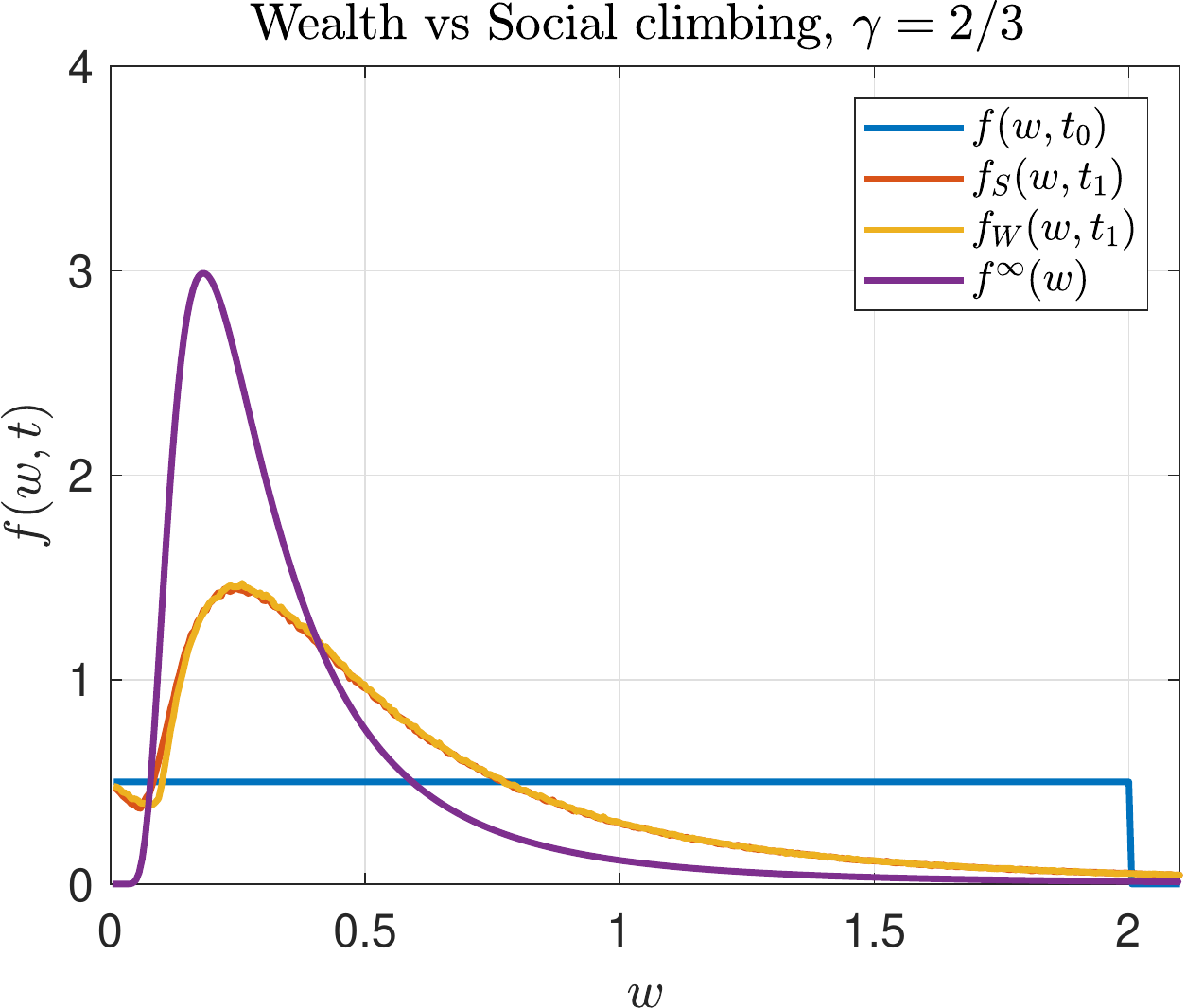}
		\includegraphics[width=5.5cm]{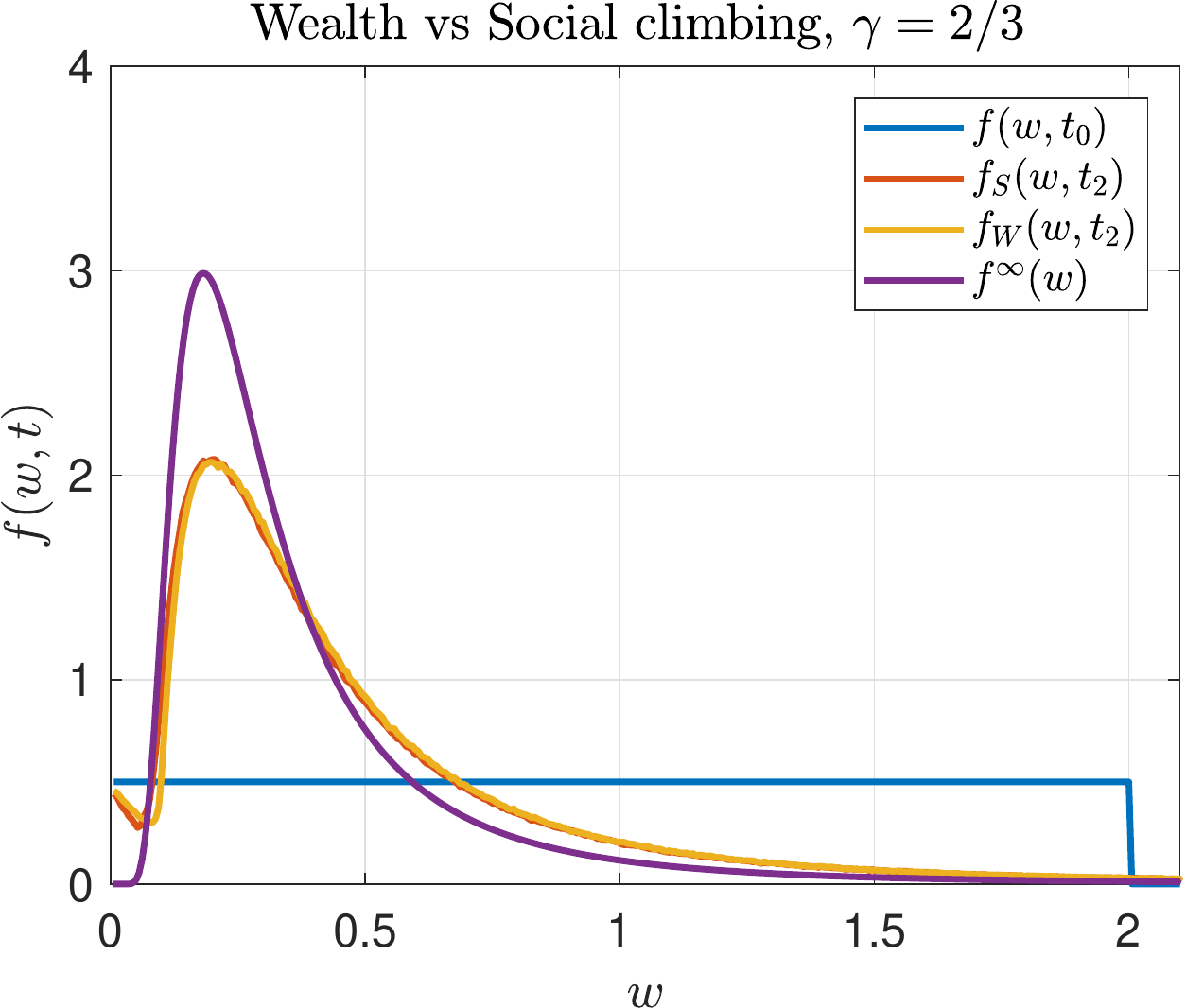}\\\vspace{1cm}
		\includegraphics[width=5.5cm]{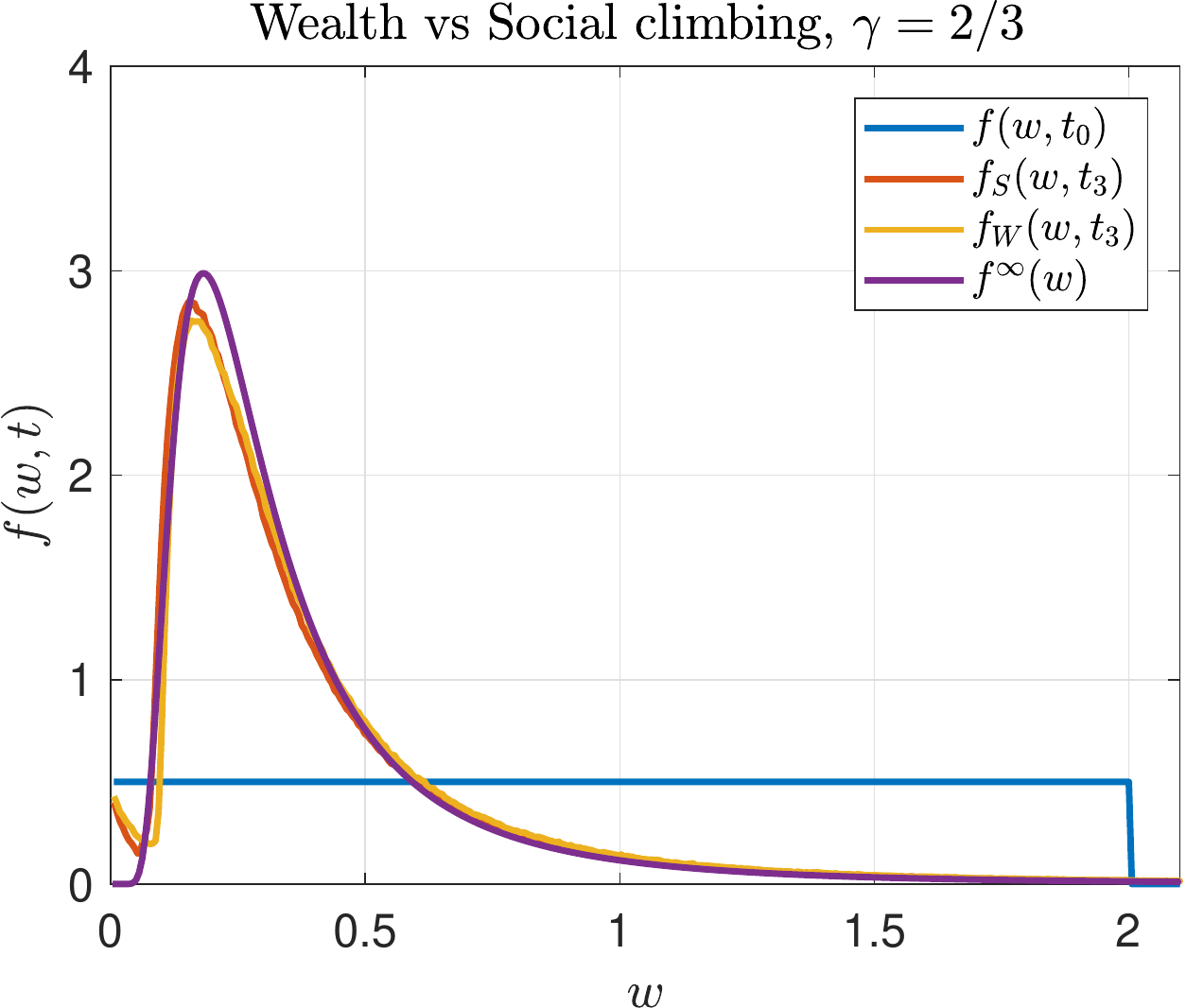}
		\includegraphics[width=5.5cm]{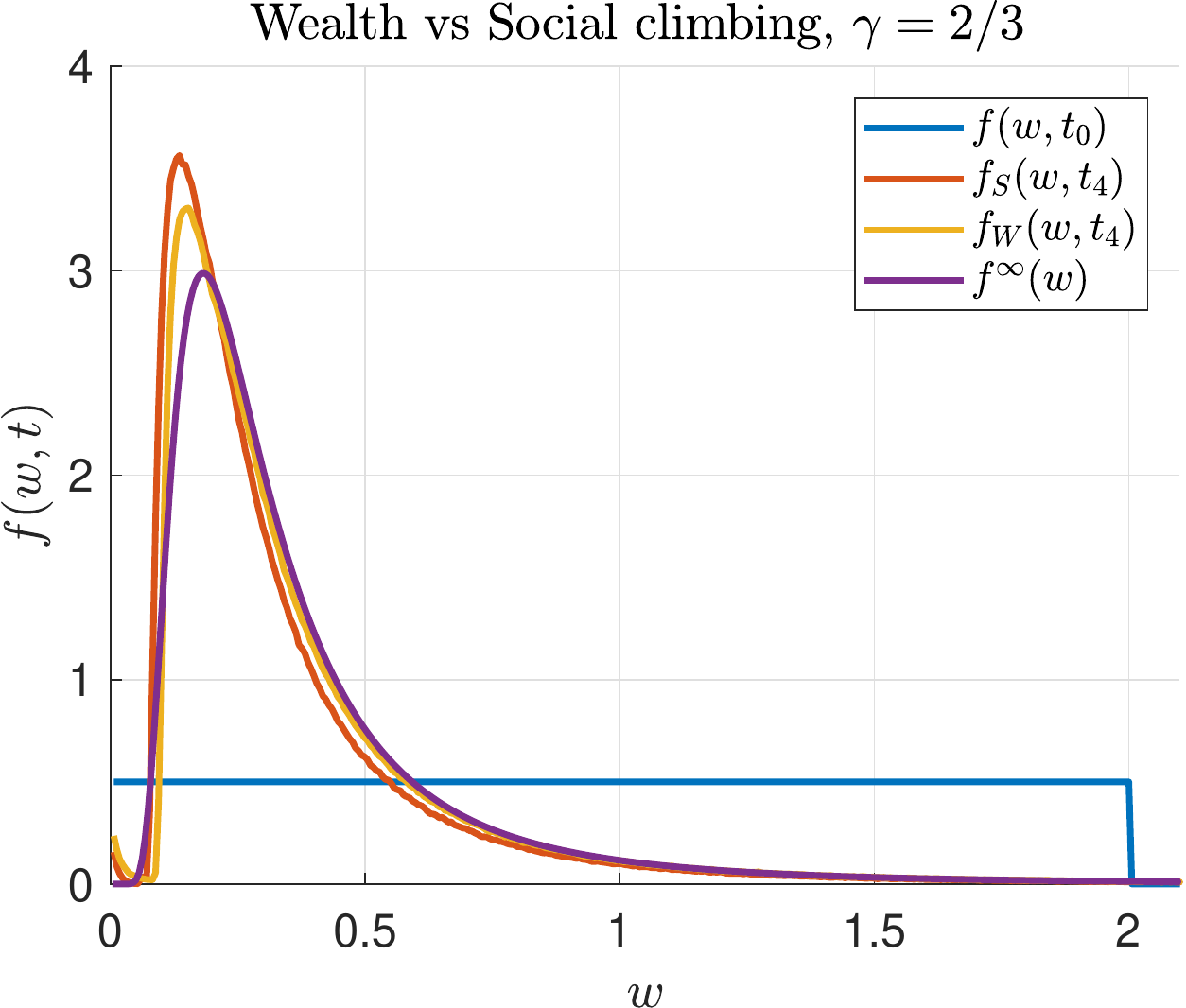}}
	\caption{Test 2. Comparison of the wealth and the social climbing models in the case $\delta=1$, $\gamma=2/3$ and $\epsilon=0.1$. From top left to bottom right the images show the trend to the steady state solution for the two models. The initial and the final Fokker-Planck states are also shown.}\label{fig:test22}
\end{figure}

\begin{figure}\centering
	{\includegraphics[width=5.5cm]{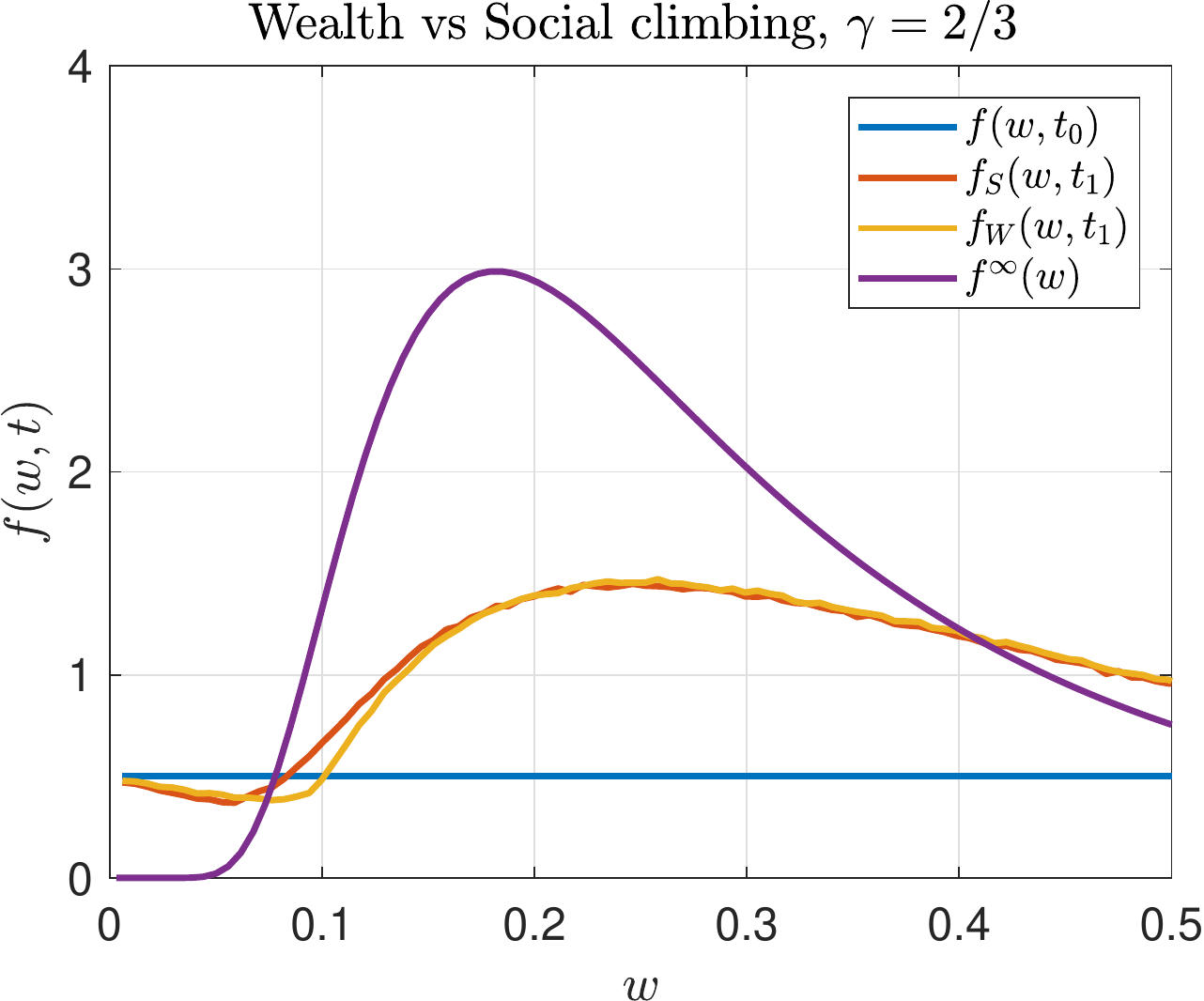}
		\includegraphics[width=5.5cm]{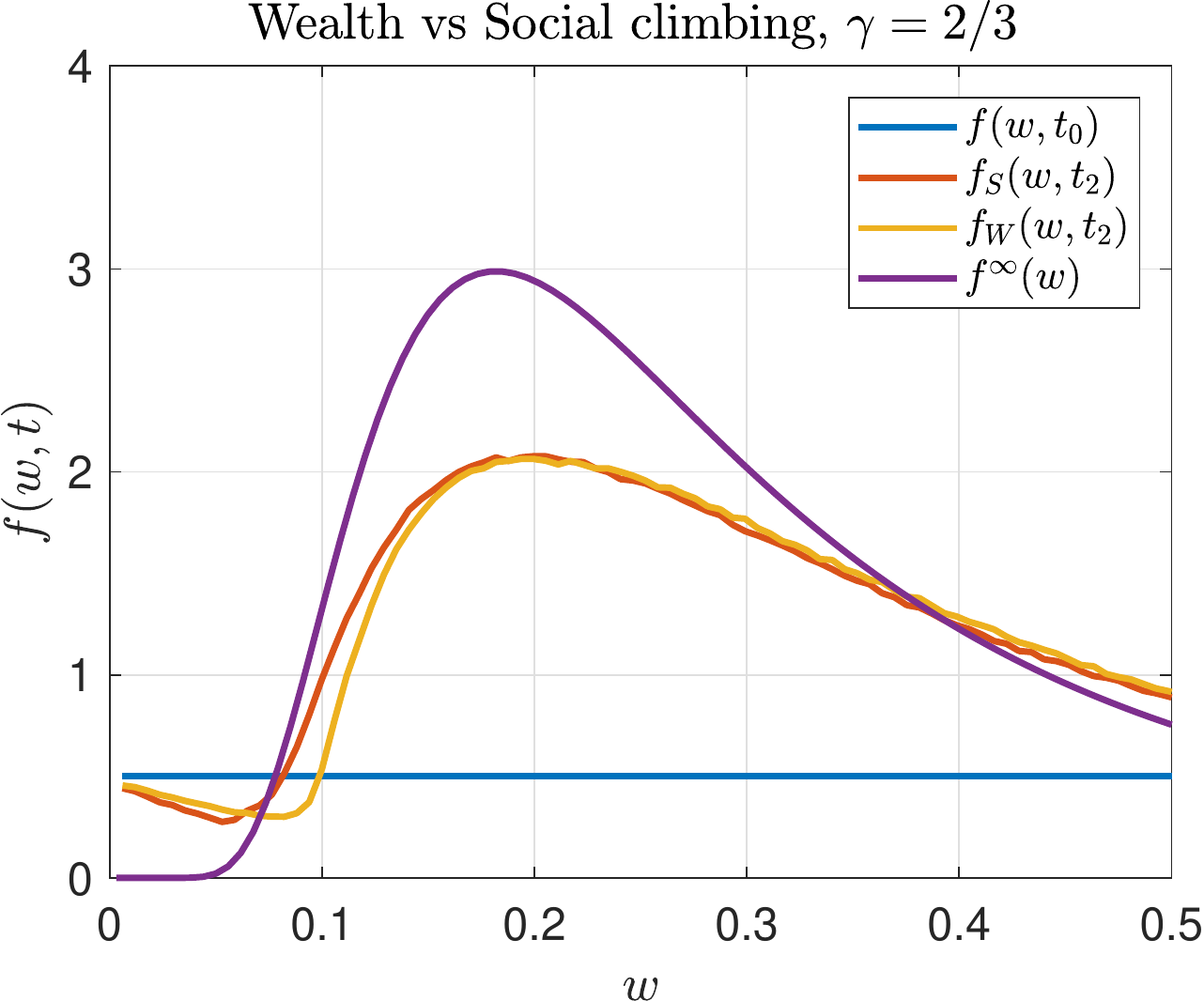}\\\vspace{1cm}
		\includegraphics[width=5.5cm]{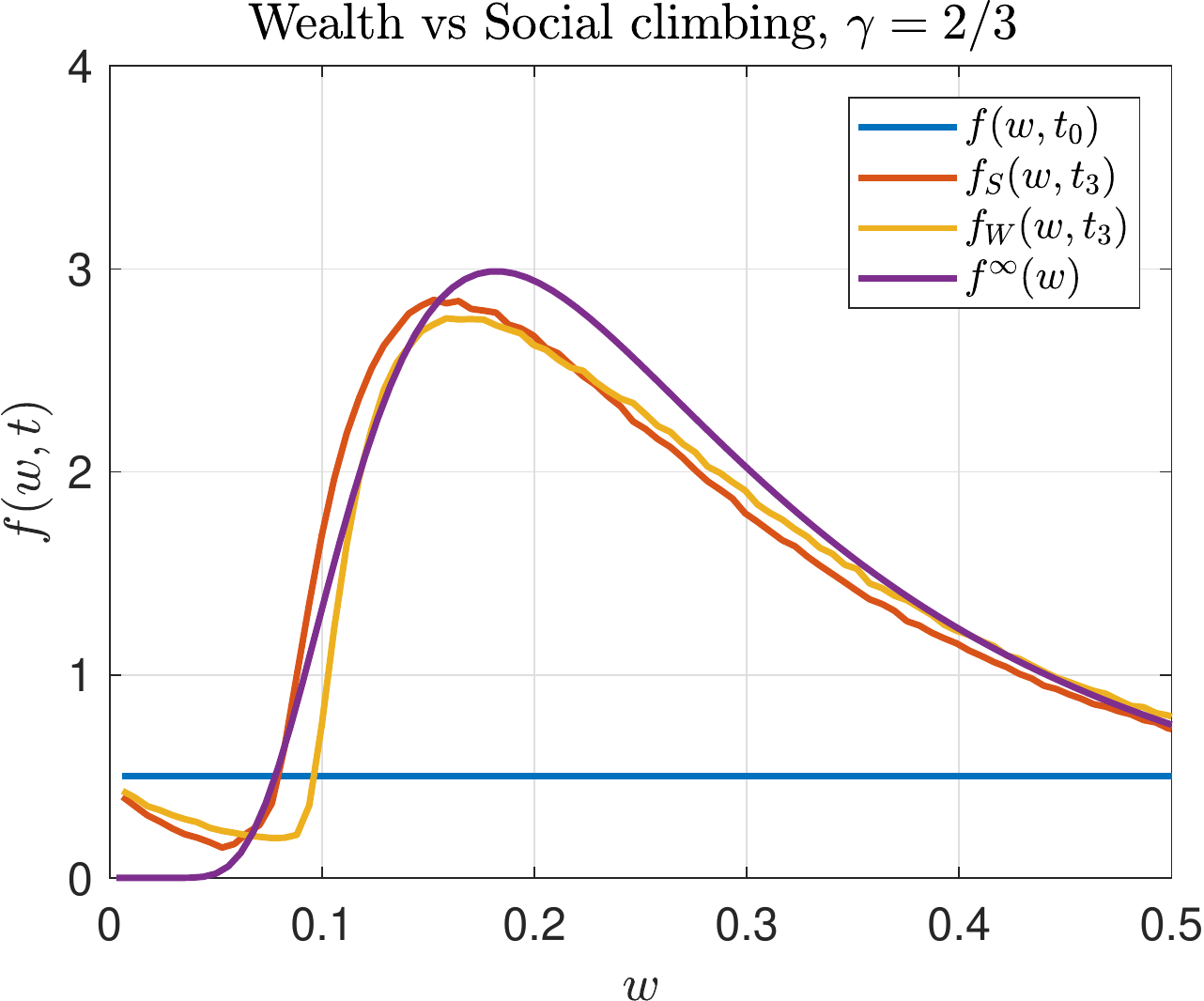}
		\includegraphics[width=5.5cm]{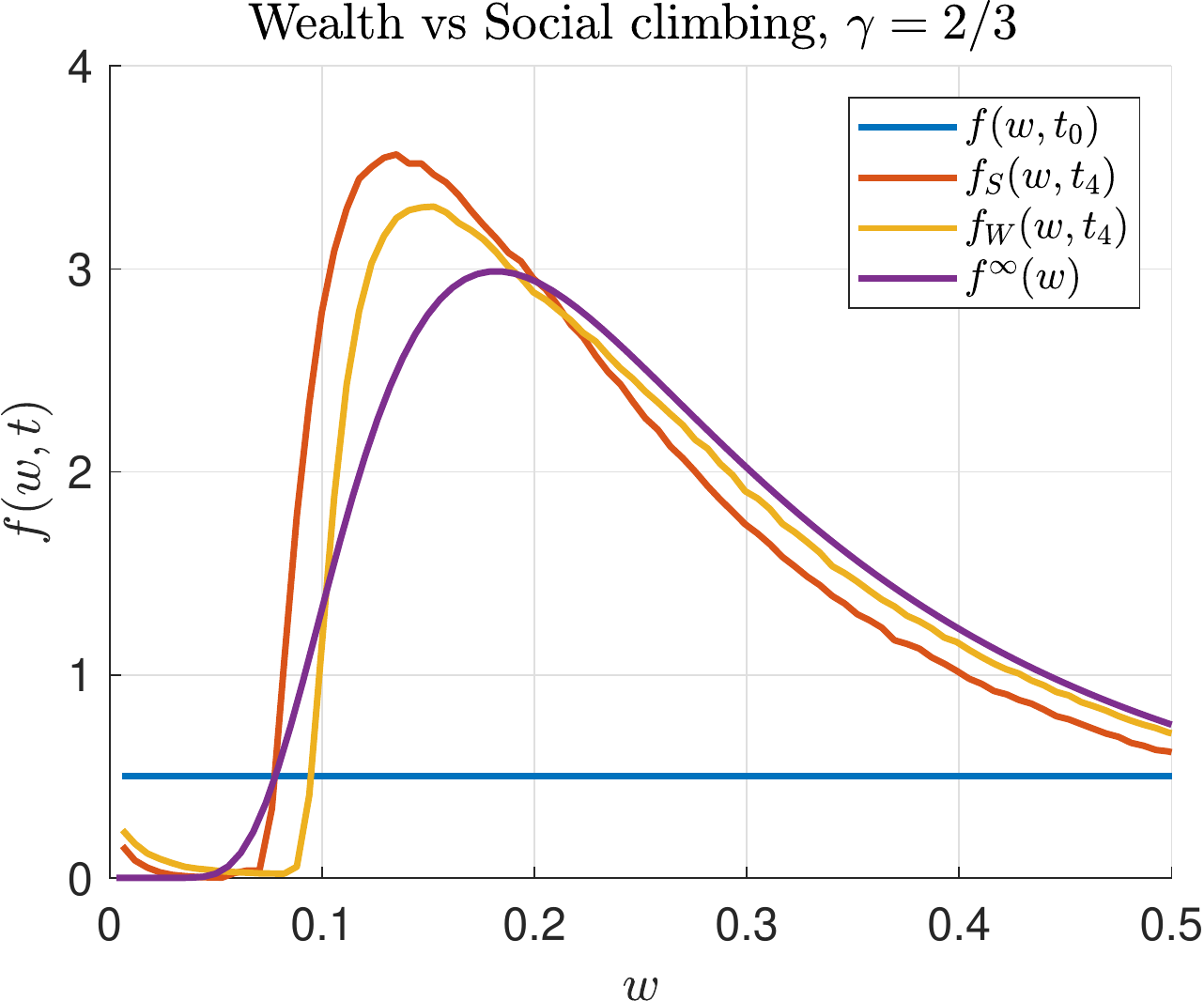}}
	\caption{Test 2. Comparison of the wealth and the social climbing models in the case $\delta=1$, $\gamma=2/3$ and $\epsilon=0.1$. From top left to bottom right the images show the trend to the steady state solution for the two models. The initial and the final Fokker-Planck states are also shown. Zoom aroung the lower wealth levels.}\label{fig:test22z}
\end{figure}

This has been done to highlight the difference between the two dynamics,  since, as shown here and in  \cite{FPTT1}, both share  the same steady state in the limit $\epsilon\to 0$ and consequently the differences in this limit setting would be very small. In Figure \ref{fig:test21z} the same curves are shown zooming around the lower wealth level where as expected the differences exhibited by the two models are larger. Each image in Figure \ref{fig:test21} and \ref{fig:test21z} reports the initial distribution, the final Fokker-Planck distribution, the distribution of the social status $f_S(t,w)$ and of the wealth $f_W(t,w)$ at different  times from top left to bottom right. The same results are shown in Figure \ref{fig:test22} and \ref{fig:test22z} where $\gamma=2/3$, the other parameters remaining unchanged. As expected, the agents belonging to the lower social class are the ones for which the two models give larger differences. 

\subsection{Wealth and Social climbing: convergence to to the Fokker-Planck dynamics and Pareto tails. }\label{numericsIII}
\begin{figure}\centering
	{\includegraphics[width=5.5cm]{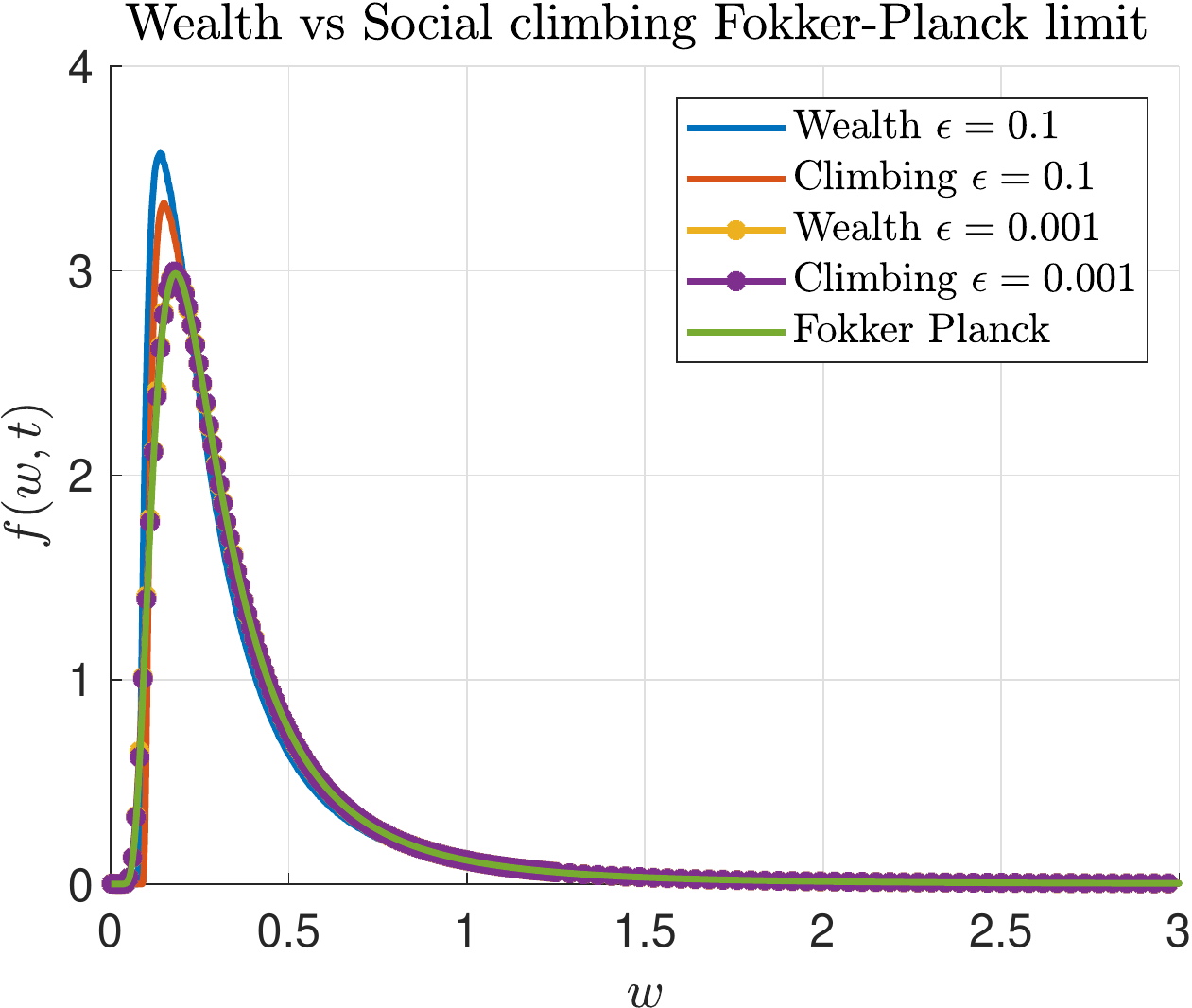}
	\includegraphics[width=5.9cm]{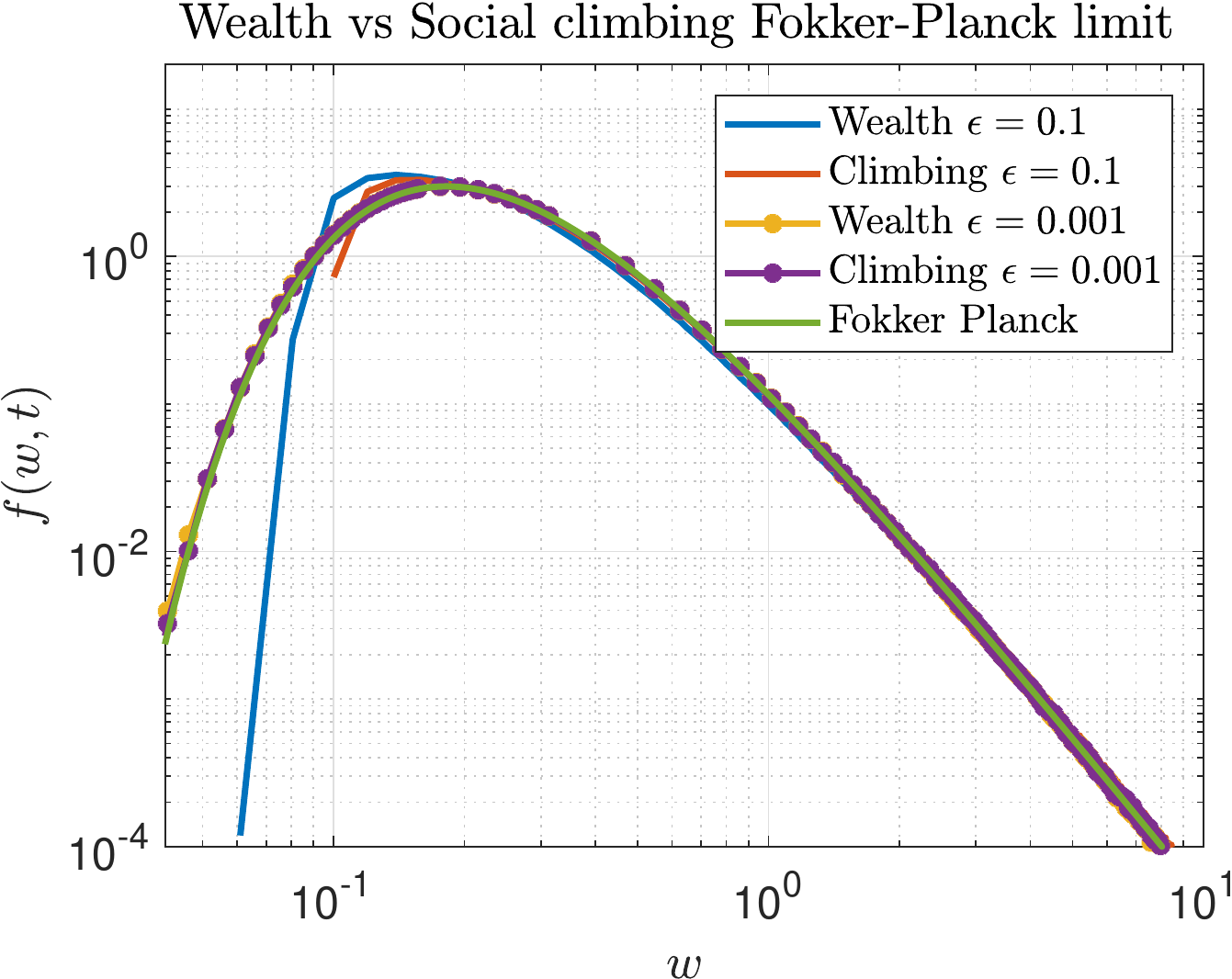}}
	\caption{Test 3. Asymptotic behavior of the Fokker-Planck model and the Boltzmann wealth and social climbing models with $\delta=0.1$, $\gamma=1$, $\bar{w}_L=1$. The right image is in log-log scale.}\label{fig:test3}
\end{figure}

In this last test, we investigate the differences between the Boltzmann and the Fokker-Planck dynamics for the social climbing interaction \eqref{vff} and the wealth model \eqref{lin2}, respectively. The final equilibrium state is in both cases represented by the same inverse Gamma distribution, i.e. $\delta=1$. This means that Pareto tails are expected to appear in this steady state limit when $\epsilon\to 0$. The Figure \ref{fig:test3} shows on the left the steady state distributions for wealth and social status for $\epsilon=0.1$ and $\epsilon=0.001$ as well as the corresponding inverse Gamma distribution \eqref{equilibrio} equilibrium state. On the right hand side, we plot the results in log-log scale, which highlights the Pareto tails index.  The different behaviors of the two models  far from the Fokker-Planck dynamics are visible, both for the agents possessing a lower wealth level as well as for the wealthy part of the population living in the tail of the distribution.
\vskip 3cm

\section{Conclusions}

The statistical distribution of social rank in a multi-agent society has been described in this paper by resorting to classical methods of collision-like kinetic theory.  The main goal of our analysis was to provide an explanation of the emergence of steady states in the form of generalized Amoroso distributions, a family of distributions with polynomial tails that represent at best the formation of a social elite. The macroscopic behavior is consequent to the choice made at the microscopic level, choice that takes into account the essential features of the human behavior related to the phenomenon of social climbing.  The kinetic modeling is similar to the one introduced in  \cite{GT17}, subsequently generalized in  \cite{DT,GT18}, in which the human behavior has been shown to be responsible of the formation of a macroscopic equilibrium in the form of probability distributions with thin tails, like the lognormal distribution or the Gamma and Weibull ones. From this point of view, the present results can be considered as an extension of the kinetic description of  \cite{DT,GT18,To2}, which allows to classify, at a microscopic level, the main differences in the elementary interaction which produce a whole class of  generalized Gamma distributions, that range from the classical Gamma density to the lognormal one.  Well-known arguments of kinetic theory allow to model these phenomena by means of a Fokker--Planck equation with variable coefficients of diffusion and drift. Interestingly enough, for this class of Fokker--Planck equations a lot of mathematical results can be proven, including the exponential convergence towards equilibrium \cite{To2}. 

A relevant part of this analysis relies in a detailed comparison of the main rules of  social climbing with thats of wealth distribution, and helps to share some light on possible improvements of the latter by taking into account in a more substantial way both the individual and social aspects of human behavior present in the former. The numerical comparison of the two models reveals indeed a marked difference between the kinetic description in the low part of the profiles, difference that disappears only in the \emph{grazing} asymptotics.


\section*{Acknowledgement} This work has been written within the
activities of GNFM group  of INdAM (National Institute of
High Mathematics), and partially supported by  MIUR project ``Optimal mass
transportation, geometrical and functional inequalities with applications''.


\end{document}